\documentclass[prx,superscriptaddress,amsmath,amssymb,twocolumn,aps,footinbib,longbibliography]{revtex4-1}

\usepackage{graphicx}
\usepackage{xcolor}
\usepackage{bm,bbm}
\usepackage{hyperref}
\usepackage{multirow}

\newcommand{\upi}{\mathrm{i}}
\newcommand{\upe}{\mathrm{e}}
\newcommand{\tr}{\mathrm{Tr}}
\newcommand{\id}{\mathbbm{1}}

\newcommand{\pthres}{\eta_\mathrm{thres}}
\newcommand{\perr}{p_\mathrm{err}}

\newcommand{\nbarg}{{\overline{n}_\mathrm{g}}}

\newcommand{\nbart}{\overline{n}_\mathrm{tot}}
\newcommand{\nbarl}{\overline{n}_\mathrm{L}}

\begin{document}

\title{Limitations in quantum computing from resource constraints}

\author{Marco \surname{Fellous-Asiani}}
\affiliation{Institut N{\'e}el, Grenoble, France}

\author{Jing Hao \surname{Chai}}
\affiliation{Centre for Quantum Technologies, National University of Singapore, Singapore}

\author{Robert S.~\surname{Whitney}}
\affiliation{Laboratoire de Physique et Mod{\'e}lisation des Milieux Condens{\'e}s,
Universit{\'e} Grenoble Alpes and CNRS, BP 166, 38042 Grenoble, France.}

\author{Alexia \surname{Auff{\`e}ves}}
\affiliation{Institut N{\'e}el, Grenoble, France}

\author{Hui Khoon \surname{Ng}}
\email{huikhoon.ng@yale-nus.edu.sg}
\affiliation{Yale-NUS College, Singapore}
\affiliation{Centre for Quantum Technologies, National University of Singapore, Singapore}
\affiliation{MajuLab, International Joint Research Unit UMI 3654, CNRS, Universit{\'e} C{\^o}te d'Azur,
Sorbonne Universit{\'e}, National University of Singapore, Nanyang Technological University, 
Singapore}

\date{\today}

\begin{abstract}
Fault-tolerant schemes can use error correction to make a quantum computation arbitrarily accurate, provided that errors per physical component are smaller than a certain threshold and independent of the computer size. However in current experiments, physical resource limitations like energy, volume or available bandwidth induce error rates that typically grow as the computer grows. Taking into account these constraints, we show that the amount of error correction can be optimized, leading to a maximum attainable computational accuracy. We find this maximum for generic situations where noise is scale-dependent.  By inverting the logic, we provide experimenters with a tool to finding the minimum resources required to run an algorithm with a given computational accuracy. When combined with a full-stack quantum computing model, this provides the basis for energetic estimates of future large-scale quantum computers.
\end{abstract}
\maketitle


\section{Introduction}\label{sec:intro}

\begin{figure*}
\includegraphics[width=0.95\textwidth]{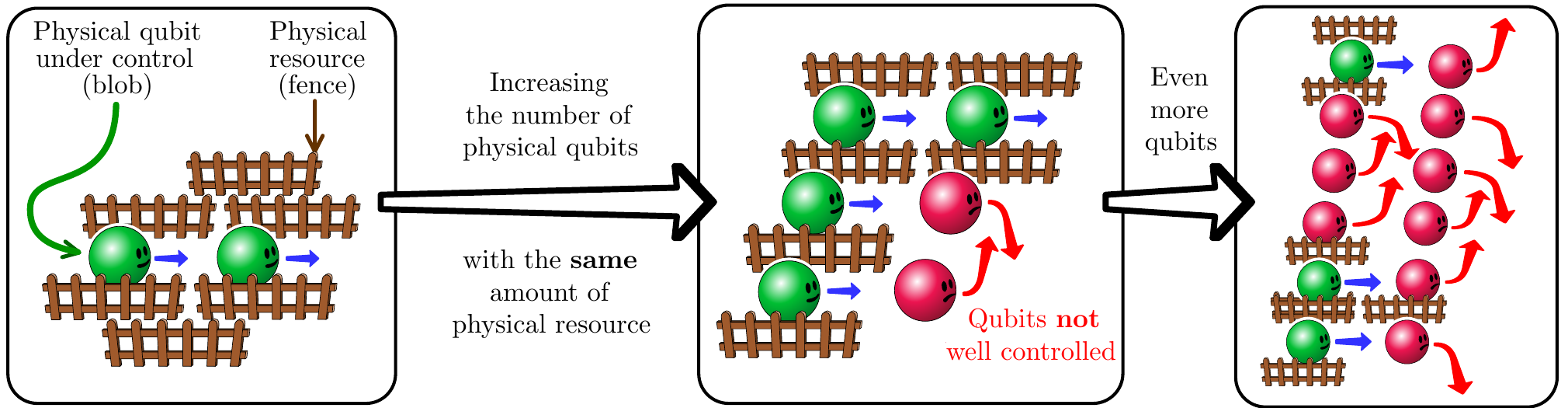}
\caption{\label{fig:Cartoon} How resource constraints can lead to increased computational errors: Each gate operation on a physical qubit (blob) requires a certain amount of physical resource (fence) for good control. If the number of physical qubits increases as the computer grows, without a proportionate increase in control resource, errors will increase.
} 
\end{figure*}

\noindent With the advent of small-scale quantum computing devices from companies like IBM, and the myriad software and hardware quantum startups, the interest in building quantum computers is at an all-time high. The latest declaration of quantum supremacy by Google \cite{Google2019} begs the question: How do we make our quantum computers more powerful? The answer is, of course, to have larger quantum computers. But larger also usually means noisier, with more fragile quantum components that can go wrong, leading to more computational errors. The way out of this conundrum is fault-tolerant quantum computation (FTQC), the only known route to scaling up quantum computers while keeping errors in check.

FTQC schemes have been known since the early days of the field \cite{Shor1996,Kitaev1997,Gottesman1997,Knill1998,Preskill1998,Aliferis2006,Aharonov2008Jul}, and are widely reviewed \cite{Gottesman2010review,NielsenChuang2010,Raussendorf2012Sep,Devitt2013Jun}.
They remain an active field of research, especially in the context of 
surface codes, see e.g., Refs.~\cite{Tuckett2020Mar,Brown2020May,BonillaAtaides2021Apr} or Ref.~\cite{Fowler2012} for an older review. Underlying all FTQC schemes are basic assumptions about the nature of the quantum devices and the noise afflicting them. Many of these assumptions, laid down long before experimental devices came about.
As we learn more about the shape of quantum computers to come, it is important to re-visit those assumptions, to update them to properly describe real devices, so that the schemes remain relevant to our progress towards large-scale, useful quantum computers.

FTQC tells us how to scale up the quantum computer, to accommodate larger problem sizes and improve computational accuracy, by increasing the physical resources spent on implementing the computation. Every known FTQC scheme relies on quantum error correction (QEC) codes to remove errors, using more and more powerful codes to remove more and more errors, accompanied by a prescription to avoid uncontrolled spread of errors as the computer grows. One key assumption is that the physical error probability $\eta$---the maximum probability that an error occurs in a physical qubit or gate---remains constant as the computer scales up.  
Then, so long as the error is below a certain threshold (typically an error probability per gate of less than $10^{-4}$), one can perform more accurate calculations by investing more physical resources to scale up the computer's size (adding more qubits, gates, etc). In principle, this can be repeated until computational errors are arbitrarily rare.

However because of resource limitations, the growth of $\eta$ with scale is observed in current quantum devices. For instance, in ion-trap experiments, the gate fidelity drops rapidly if more and more ions are put into the same trap; this volume constraint is the motivation behind the push for networked ion traps and flying qubits to communicate between traps (see, for example, \cite{Monroe2013}). Another example is provided by qubits that are coherently controlled, by resonantly addressing their transition. Here a constraint on the total available energy to perform gates results in lower gate fidelity\cite{Miko}. Finally, a constraint on the available bandwidth makes the qubit transition frequencies closer and closer as the computer size grows, causing more and more crosstalk between qubits when performing gates \cite{Google2019}. These three typical examples lead to a scale-dependent noise.

If the physical error probability $\eta$ is scale-dependent, so that it grows as the computer scales up, we cannot expect quantum error correction to keep up with the rapid accumulation of errors, so it should come as no surprise that the standard threshold for fault-tolerance \cite{Shor1996,Kitaev1997,Knill1998,Preskill1998,Aharonov2008Jul,Gottesman1997,Aliferis2006}---and its generalizations to correlated and long-range noise \cite{Terhal2005Jan,Aharonov2006Feb,Novais2006Jul,Novais2008Jul,Aliferis2008Nov,Ng2009Mar,Preskill2013Mar,Novais2013Jan,Fowler2014Mar,Jouzdani2014Oct}---no longer applies (see e.g.~Ref.~\cite{Aharonov2006Feb,gea2002some}). This calls to revisit the expectations with respect to FTQC within this realistic context.

In this work, we examine the consequences on FTQC of growing physical error probability $\eta$ as the computer scales up. 
The absence of a threshold means that arbitrarily accurate computation is unattainable, but it does not mean that quantum error correction is useless. We find generic situations where a certain amount of error correction is good, but too much is bad.  Hence,
the amount of error correction should be optimized, leading to a maximal achievable computational accuracy. We provide experimenters with a methodology to estimate this maximum for a given scale dependent noise, and show the importance of adjusting the experimental design to control this scaling. Inverting the perspective allows us to estimate the minimum required resource cost to perform a computation with a given accuracy. 

After recalling the basics of FTQC, we present our general strategy. We exemplify it with a toy model that captures the main features of FTQC in the presence of scale dependent noise. We then focus on three physically motivated situations where resource constraints like energy, volume or bandwidth lead to scale dependent noise, and exam the feasibility of FTQC in the limit of large quantum computers. We finally provide first methodological steps towards minimizing the energetic costs to run an algorithm with a given accuracy. This suggests the possibility of a detailed energetic analysis for a full-stack quantum computer, which however goes beyond the scope of this paper.

\section{ACCURATE QUANTUM COMPUTING}\label{sec:FTIntro}

To be concrete, we examine the FTQC scheme of Ref.~\cite{Aliferis2006}, built on the idea of concatenating a QEC code put forth in earlier works. This formed the foundation of many subsequent FTQC proposals; our results are hence applicable to those based on concatenating codes. Such schemes have more well-established and complete theoretical analyses than some of the more recent developments like surface codes. They are hence a good starting point for our investigation here.

Universal quantum computation in the scheme of \cite{Aliferis2006} is built upon the 7-qubit code \cite{Steane1996}, using seven physical qubits to encode one (logical) qubit of information. We refer to the seven physical qubits used to encode the logical qubit as a ``code block", and gates on the logical qubit as ``encoded gates". At the lowest level of protection against errors, which we refer to as ``level-1 concatenation", each logical qubit is encoded using the 7-qubit code into one code block, and every computational gate is done as an encoded gate on the code blocks. Every encoded gate is immediately followed by a QEC box, comprising syndrome measurements to (attempt to) correct errors in the preceding gate. Faults can occur in any of the physical components---physical qubits and gates---including those in the QEC boxes, so the error correction may not always successfully remove the errors. Faulty components in the QEC box may even add errors to the computer. A critical part of the construction of Ref.~\cite{Aliferis2006} is to ensure that the QEC boxes, even when faulty, do not cause or spread errors on the physical qubits in an uncontrolled manner provided not too many faults occur, a realization of the notion of fault tolerance.

\begin{figure}
\includegraphics[width=0.85\columnwidth]{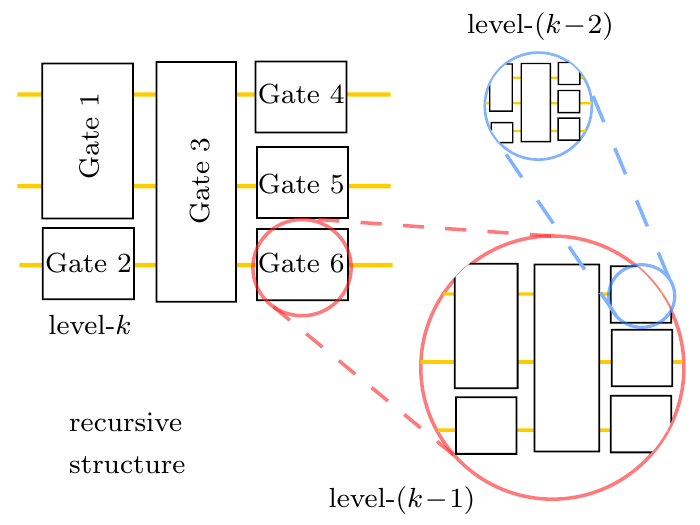}
\caption{\label{fig:recur} The accuracy of a quantum computation can be increased by a FTQC scheme that makes use of concatenation and recursive simulation. Circuits are designed to be hierarchical, with high-level gate components built from lower-level components in a self-similar manner.} 
\end{figure}

At level-1 concatenation, the ability of the code to remove errors is limited. The 7-qubit code ideally removes errors in at most one of the seven physical qubits in the code block. To increase the QEC power, we raise the concatenation level of the circuit: Every physical qubit in the lower concatenation level is encoded into seven physical qubits; every physical gate is replaced by its 7-physical-qubit encoded version, followed by an QEC box. In this manner, level-$k$ concatenation is promoted to level-$(k+1)$ concatenation, for $k=0,1,2,\ldots$. The QEC ability of each level of concatenation increases in a hierarchical manner. For example, at level-2 concatenation, every logical qubit is stored in $7^2=49$ physical qubits organized into two layers of protection, with the topmost layer comprising seven blocks of seven physical qubits each. Each block of seven physical qubits is protected using the 7-qubit code; the 7 blocks of qubits are themselves protected by QEC in the second layer. This logic extends to higher levels of concatenation. 

The concatenation endows the overall computational circuit with a recursive structure (see Fig.~\ref{fig:recur}), a crucial ingredient in the proof of the quantum accuracy threshold theorem. The increase in computational resource as the concatenation level grows is beneficial only if the increased noise due to the larger circuit is less than the increased ability to remove errors. This leads to the concept of a fault-tolerance threshold condition. The quantum accuracy threshold theorem gives a prescription for increasing the accuracy of quantum computation with no more than a polynomial increase in resources, provided the physical error probability is below a threshold level. Specifically, for the FTQC scheme of \cite{Aliferis2006}, the error probability per \emph{logical} gate at level-$k$ concatenation is upper-bounded by 
\begin{equation}\label{eq:pk}
p^{(k)}=\frac{1}{B}(B \eta)^{2^k}.
\end{equation}
Here, $\eta$ is the \emph{physical} error probability, and $p^{(0)}=\eta$. $B$ is a numerical constant, determined by the fault tolerance scheme, that captures the increase in complexity (number of physical components) of the circuit used to implement a single logical gate as one increases $k$ for increased protection. Eq.~\eqref{eq:pk} expresses quantitatively the idea of the accuracy threshold theorem: As long as
\begin{equation}\label{eq:thres}
\eta <\frac{1}{B}\equiv \eta_\mathrm{thres}, 
\end{equation}
$p^{(k)}$ decreases as $k$ increases. Eq.~\eqref{eq:thres} is the threshold condition, i.e., the physical error probability $\eta$ in the quantum computer has to be below the threshold level $\eta_\mathrm{thres}$ for FTQC to work. 
The number of physical gates in the circuit that implements the level-$k$ logical gate is $G^{(k)}= A\, (A')^{k-1}$,
where $A'$ and $A$ are integers given by circuit details;
the well-known scheme of Ref.~\cite{Aliferis2006} has $A=575$ and $A'=291$ \footnote{Both $A$ and $A'$ are independent of the nature of the errors, and so independent of the system size.  
The integer $A'=291$ is the number of physical gate operations in a CNOT's ``Rec'', and $A=575$ is the number in its ``exRec''.  
''Rec'' is the encoded CNOT plus the following QEC box, while ``exRec'' is the encoded CNOT plus the preceding and following QEC boxes, see Ref.~\cite{Aliferis2006}.}. From $A$ and $A'$, one counts the number of fault locations, $B$. A simple over-estimate of the integer $B$ is $\binom{A}{2}\simeq 10^{5}$, with a more careful counting giving an improved value of $B\simeq 10^4$ \cite{Aliferis2006}.

The quantum accuracy threshold theorem \cite{Shor1996,Kitaev1997,Knill1998,Preskill1998,Aharonov2008Jul,Gottesman1997,Aliferis2006} shows that a double-exponential decrease in $p^{(k)}$ with $k$ can be achieved  with only an exponential increase in resources, giving the no-more-than-polynomial increase in resource costs.
This quantum accuracy threshold theorem assumes that the value of $\eta$, the physical error probability, remains constant even as the level of concatenation $k$ increases. However, as mentioned, as $k$ increases and the physical size of the computer grows, current experiments suggest that $\eta$ also increases. Our goal here is thus to examine how the conclusions on quantum accuracy are modified if $\eta$ grows with $k$. Then it is intuitively clear that as $k$ is increased in an attempt to reduce the logical error probability, the underlying noise per physical component increases to thwart that reduction. We will see that there is a maximum $k$ beyond which further concatenation only serves to worsen the computational accuracy.


\section{Effect of scale-dependent noise}\label{sec:scaleDepErr}
\begin{figure}
\includegraphics[width=0.85\columnwidth]{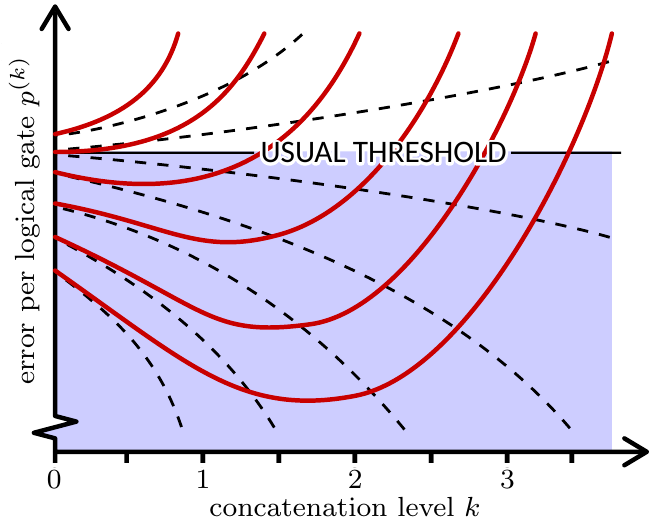}
\caption{\label{fig:Thres} 
The black dotted curves are a schematic of the conventional situation (where the physical error probability $\eta$ is independent of the scale---the concatenation level $k$--- of the computer. The red solid curves are a schematic of this work's consideration, where $\eta$ grows with $k$ (each red curves is for a different value of $p^{(0)}\equiv\eta$). If $\eta$ is $k$ independent, standard FTQC analysis says that the error per logical gate $p^{(k)}$ can be brought as close to $0$ as desired by increasing $k$, provided one starts below the threshold (solid horizontal line) at $k=0$. If $\eta$ depends on $k$, even if one starts below the threshold, $p^{(k)}$ eventually turns around for large enough $k$; $p^{(k)}$ cannot reach $0$, there is a maximum concatenation level, and further increase in $k$ only increases the logical error. All examples in this work have at most one minimum in each red curve, but in general a curve can be multiple minima.
} 

\end{figure}

We examine the consequences of a $k$-dependent physical error probability $\eta^{(k)}$, illustrating it first with a toy model, before analyzing the more realistic situation where a constraint on the total resource available for the computation leads to a shrinking amount of resource per physical gate as the computer scales up. The general effect of a $k$-dependent physical error probability can be summarized in the schematic Fig.~\ref{fig:Thres}. We also show how to obtain the maximum computational accuracy available for a given model for scale-dependent error.

\subsection{Toy model}\label{subsec:toy}

\begin{figure}
\includegraphics[width=0.94\columnwidth]{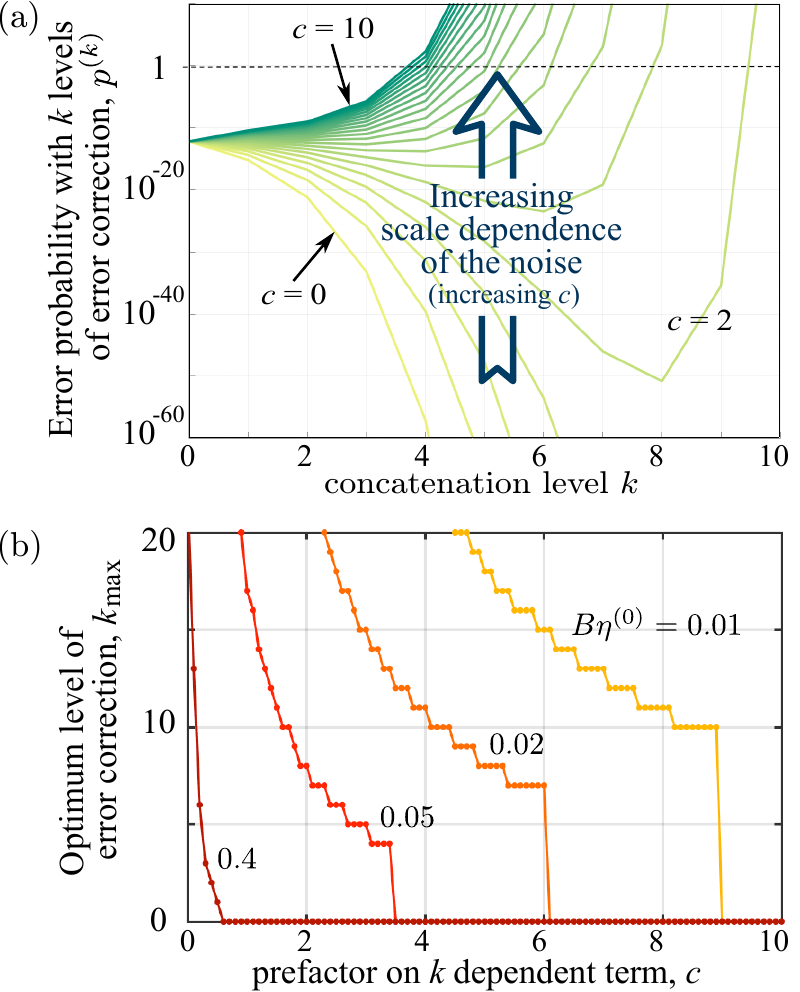}
\caption{\label{fig:Toy} (a) An example of how $p^{(k)}$ varies as the concatenation level $k$ increases, for the affine model with $c=0$ (lightest color), $0.5, 1, \ldots, 10$ (darkest color), $B=10^4$ and $\eta^{(0)}=5\times 10^{-6}$, for which $B\eta^{(0)}=0.05$, far below the threshold for $c=0$. All curves, apart from that for $c=0$, turn around for large enough $k$. (b) Maximum $k$ value, $k_{\max}$, such that $p^{(k+1)}<p^{(k)}$, with $p^{(0)}\equiv \eta^{(0)}$, the unencoded error probability, for different values of $B\eta^{(0)}<1$ and $c\in\{0, 0.1, 0.2,\ldots,10\}$. In every case, $k_{\max}$ eventually falls to zero for large enough $c$, i.e., it is better to have no encoding. 
}
\end{figure}

We first illustrate this with a simple model in which $\eta^{(k)}= \eta^{(0)}(1+ck)$, for $k=0,1,2,\ldots$, where $\eta^{(k)}$ is the physical error probability per gate  in a computer large enough to perform level-$k$ concatenation. Here, $c\geq 0$ and $\eta^{(0)}\geq 0$ are constants governed by the physical system in question. Although this is a toy model, one can think of it as the affine approximation of any $\eta^{(k)}$ function with weak $k$ dependence, expanded about $\eta^{(k)}=\eta^{(0)}$ (see also the long-range noise with $z=d$ in Table~\ref{tab:scale_dep_noise}).  For this $\eta^{(k)}$, Eq.~\eqref{eq:pk} gives
\begin{equation}
p^{(k)}=\frac{1}{B}{\left[B\eta^{(0)}(1+ck)\right]}^{2^k}=p_0^{(k)}(1+ck)^{2^k}. 
\label{Eq:linear-approx}
\end{equation}
Here, we define 
\begin{equation}
p_0^{(k)}\equiv \frac{1}{B}{\left(B\eta^{(0)}\right)}^{2^k},
\label{Eq:definition-p0^(k)}
\end{equation} 
which would be the value of the error probability per logical gate if the error probability per physical gate were $k$ independent. If $c=0$, $p^{(k)}/p^{(k-1)}=p_0^{(k)}/p_0^{(k-1)}<1$ as long as $\eta^{(0)}<1/B$, as in Eq.~\eqref{eq:thres}; if $c>0$, the multiplicative factor $(1+ck)$ grows with $k$ so, eventually, $p^{(k+1)}>p^{(k)}$ for $k$ beyond some $k_{\max}$ value. Figure \ref{fig:Toy}(a) shows an example of how $p^{(k)}$ varies as $k$ increases, for different $c$ values. As long as $c>0$, $p^{(k)}$ decreases (if at all) before rising again, above some $k_{\max}$ value. 

Corresponding to this maximum useful level of concatenation $k_{\max}$ is the minimum attainable error probability, $p_{\min}\equiv p^{(k_{\max})}$, giving the limit to computational accuracy attainable for given values of $c$ and $B\eta^{(0)}$, quantities that give information about the noise scaling and the fault-tolerance overheads.
Figure \ref{fig:Toy}(b) shows the $k_{\max}$ values for different $c$ and $B\eta^{(0)}$ values (c.f.~Fig.~\ref{fig:Thres}b). Clearly, $k_{\max}$ decreases as $c$ grows (stronger $k$ dependence). Current experiments have $B\eta^{(0)}\gtrsim 1$; for example, the IBM Quantum Experience system has $\eta^{(0)}\gtrsim 10^{-3}$, giving $B\eta^{(0)}\gtrsim 10$ for $B= 10^4$. In near- to middle-term experiments , we expect $B\eta^{(0)}$ to not be far below 1, i.e., the error probability is just below the $c=0$ threshold value [see Eq.~\eqref{eq:thres}]. In this case, Fig.~\ref{fig:Toy}b suggests that one quickly loses the advantage of concatenating to higher levels even for small $c$ values. In fact, for encoding to be helpful at all, i.e., for $k=1$, we must have $p^{(1)}=B[\eta^{(0)}]^2(1+c)^2<p^{(0)}=\eta^{(0)}$, which in turn requires
\begin{equation}\label{ccond}
c<\frac{1}{\sqrt{B\eta^{(0)}}}-1.
\end{equation}
If $B\eta^{(0)}=0.8$, say, this amounts to the requirement that $c\lesssim 0.1$, so that a very weak dependence on $k$ is necessary for even one level of encoding to help at all in reducing the error probability.

\subsection{General case}\label{subsec:Gen}
Consider the physical error probability growing as a monotonic function of $k$: $\eta^{(k)}=\eta^{(0)}f(k)$, with $f(k)\geq 1$ monotonically growing with $k$, and $f(0)=1$. Then,
the error probability per logical gate is $p^{(k)}=p_0^{(k)}f(k)^{2^k}$, where Eq.~(\ref{Eq:definition-p0^(k)}) gives $p_0^{(k)}$. Treating $k$ as a continuous variable,let us assume there is only one minimum, which we define as $k=k_{\rm st}$ (with ``st'' for stationary point).
Then the minimal attainable error will occur at an integer $k_{\rm max}$, which is one of the two integers 
nearest to the minima $k_{\rm st}$, so the minimal error will occur at $k_{\max}\leq k_{\rm st}+1$. 
If there are multiple minima, we define $k_{\rm st}$ as the minimum with the largest $k$. A priori, we do not know which minimum will be the best, but we still know that $k_{\max}\leq k_{\rm st}+1$.
Combining this with a little algebra for $k_{\rm st}$ yields 
\begin{equation}
k_{\max}\,< \,1+ f^{-1}\!{\left(\frac{1}{B\eta_0}\right)},
\end{equation}
with $f^{-1}(\cdot)$ the inverse of $f(\cdot)$.  
 If $f(k\!\to\!\infty)$ is finite, $p^{(k)}$ can be made arbitrarily small only if $\eta^{(0)}<[Bf(k\!\to\!\infty)]^{-1}$. However in many cases (such as the above toy model and the example in the following section), one has $f(k\!\to\! \infty) \!\to\! \infty$. Then the minimum $p^{(k)}$ will occur at finite $k_{\rm max}$, no matter how small $\eta^{(0)}$ is; one can never attain arbitrarily small logical error probability by concatenating further.


\section{Examples of Resource constraints}

We now give examples of how specific physical resource constraints can lead to the scale-dependent noise discussed in the previous section. In the first example, the constraints lead to a scale-dependent {\it local noise} on each physical component,  so the above theory applies directly.
In the second example, the constraints lead to scale-dependent {\it crosstalk} between qubits, which can be mapped to the above theory, using a mapping in Refs.~\cite{Aharonov2006Feb,Preskill2013Mar}.

\subsection{Resource constraints affecting local  noise}\label{subsec:ResConstr}

This section considers the local noise on each qubit $\eta^{(k)}$ scaling with a total number of physical components 
(gates, qubits, or similar) $N(k)$ which grows exponentially with $k$. 
Let us assume that adding a level of concatenation involves replacing each physical component by $D$ physical components, so $N(k) = D^k$. 
For the noise, we take $\eta^{(k)}\propto N(k)^\beta$ for some positive constant exponent $\beta$, so
\begin{equation}
\eta^{(k)}=\eta^{(0)}D^{\beta k}.
\label{Eq:expon}
\end{equation}
There could be various origins for such a scaling, however a common one will be total resource constraints.
One expects the resources needed to maintain a given quality of physical gate operations to scale with $N$, so a constraint on the \emph{total} available resource will result in a fall in the resource per physical component as the computer scales up. This gives a consequential drop in the quality of the gate, or, equivalently, a rise in the physical error probability $\eta$. 

The error probability per logical gate is then $p^{(k)}=p_0^{(k)} D^{\beta 2^k k}$, where Eq.~(\ref{Eq:definition-p0^(k)}) gives $p_0^{(k)}$.
Going from $(k-1)$ to $k$ levels of concatenation reduces the logical error probability when $p^{(k)}\big/p^{(k-1)} <1$, this is only satisfied for the model of Eq.~(\ref{Eq:expon}) when
$k\leq k_{\rm max}$,
where $k_{\rm max}$ is the largest positive integer satisfying, see Appendix~\ref{app:p_kmax},
\begin{equation}
k_{\rm max} < - \ln {\left(B\eta^{(0)}D^\beta\right)}\big/ \ln{\left(D^\beta\right)}.
\label{Eq:kmax-bound}
\end{equation}
If no positive integer $k_{\rm max}$ satisfies this inequality, then $k_{\rm max}=0$
and concatenation is not useful at all.  This is because concatenation is only useful if
 $p^{(1)}<\eta^{(0)}$, which requires 
\begin{equation}
\eta^{(0)} <B^{-1}D^{-2\beta}.
\label{Eq:1-level-correction}
\end{equation}
This is often a much more stringent condition than $\eta^{(0)}< \pthres$ in Eq.~(\ref{eq:thres}).
For example, if the noise scales with the number of gates in a concatenated FTQC scheme, 
we can take $N(k)=G^{(k)}\equiv A^k$ (see paragraph following Eq.~\eqref{eq:thres} above). This means we set
$D=A'$ where $A'=291$ as in Ref.~\cite{Aliferis2006}, then one sees for $\beta=1$  that
 concatenation is only useful if $\eta^{(0)} <B^{-1}A^{-2\beta} \sim 10^{-9}$ which is  $10^5$ times smaller than the usual threshold, $\pthres$.
 
 This condition is so stringent because $D$ is so large.
If the noise scales with a different physical parameter (number of qubits, number of wires, or similar), the value of $D$ will be different but it will typically still be large.
 Eq.~(\ref{Eq:1-level-correction}) then makes it clear that the larger a given parameter's  $D$ is, the more important it is to minimize the noise's scaling with that parameter (i.e. to minimize $\beta$).

The minimal attainable error probability per logical gate is given by taking $k=k_{\rm max}$
in the above formula for $p^{(k)}$.
For fixed system parameters ($D,B,\eta_0,\beta$), this   $p^{(k_{\rm max})}$
is easily found by taking $p^{(k)}$ for different integer $k$s to see which is smallest.
However, to see its dependence on those parameters, Appendix~\ref{app:p_kmax} gives algebraic formulas for upper and lower bounds on  $p^{(k_{\rm max})}$.


\subsection{Resource constraints affecting crosstalk}
\label{sec:long-range}

A common problem in existing prototype quantum computers is crosstalk between qubits. This is an example of a more general problem of non-local non-Markovian noise, usually called long-range correlated noise. 
To treat this, we 
follow Refs.~\cite{Aharonov2006Feb,Preskill2013Mar}, and define $H_{ij}$ as the arbitrary (and potentially noisy) 
unwanted interaction between physical qubits $i$ and $j$.
This interaction could be direct, or it could be mediated by other degrees of freedom (which one traces out). In the latter case, it could be non-Markovian, meaning it can account for interactions mediated by sub-Ohmic, Ohmic or super-Ohmic baths \footnote{One simply requires that the bath spectrum has cut-offs that ensure that $||H_{ij}||$ is not divergent.}.
One then defines the error strength 
\begin{eqnarray}
\Delta = \max_i \bigg(\sum_{j=1}^N ||H_{ij}|| \bigg)
\label{Eq:Aharonov-Kitaev-Preskill}
\end{eqnarray}
for  a computer containing $N$ physical qubits.
Refs.~\cite{Aharonov2006Feb,Preskill2013Mar} showed 
that $t_0\Delta$ is a good measure of the error per gate,
where $t_0$ is the duration of the slowest physical gate, although it should not be interpreted directly as the error probability per gate, see e.g., Ref.~\cite{Aliferis2013Sep}. They then showed
that fault-tolerance occurs when $t_0 \Delta < (2e^{2+1/e}B^2)^{-1} \sim 10^{-9}$ for $N\to \infty$.

Here, in contrast, we consider cases where $\Delta$ diverges for $N\to \infty$, violating the condition for fault-tolerance in  Refs.~\cite{Aharonov2006Feb,Preskill2013Mar}. This growth of $\Delta$ with $N$ will often occur due to resource constraints.
A simple example would be a constraint on the physical volume of quantum computer, which is a current limitation in qubit technologies based on ion traps \footnote{More generally, volume constraints  will naturally occur for any technology requiring two-qubit gates between arbitrary qubits, if such gates become inaccurate at long distances.}.  Then the density of qubits must scale like $N$. If each qubit has unwanted interactions with all other qubits within a given radius, one would have $\Delta \propto N$.
A second example---relevant to multiple technologies---is a bandwidth constraint, i.e. a limit on the  available range of transition frequencies of the qubits. This is particularly a problem for existing superconducting and ion-trap qubit technologies. There each two-qubit gate corresponds to a different transition frequency, and a given gate is performed by sending a driving signal (typically a microwave signal) into the quantum computer with the frequency of that gate.
This means that the driving signal for a two-qubit gate between qubits $n$ and $m$ will also cause an unwanted interaction $H_{ij}$ for
pairs of qubits $i$ and $j$ with frequencies too close to $n$ and $m$ \footnote{This mechanism would also mean that single qubit gates cause local noise on other qubits with nearby frequencies. This has an effect like in Sec.~\ref{subsec:ResConstr}, so we do not consider it further here, and focus on the effect of $H_{ij}$.}. In some technologies, all qubits feel the driving signal, then the number of qubits feeling this unwanted interaction grows with the number of qubits in any given window of transition frequencies, which grows like $N$.  
In this case $\Delta \propto N$, however clever engineering may well reduce $\Delta$'s scaling with $N$, so we prefer to consider $\Delta \propto N^\beta$ with arbitrary $\beta$  \footnote{For example, if driving signals do not affect all qubits equally, machine learning can be used to choose each qubit frequency to minimizes crosstalk \cite{Klimov2020Jun}, thereby reducing $\beta$.}.

Now we study how the physics depends on the scaling of $\Delta$ with $N$. By taking $\Delta \propto N^\beta$, we have $\Delta \propto D^{\beta k}$ for $k$ levels of concatenation, where the number of physical qubits increases by a factor of $D$ with each level of concatenated error correction.  
We then use the method of
Refs.~\cite{Aharonov2006Feb,Preskill2013Mar}, which involves taking all results in Sec.~\ref{sec:scaleDepErr} above and replacing $\eta^{(k)}$ by $e^{1+1/(2e)}\sqrt{2 t_0\Delta}$, where $t_0\Delta  \propto D^{\beta k}$ (see Appendix~\ref{app:long_range}).  
Defining $\Delta_{\rm L}^{(k)}$ as the upper bound on effective long-range correlated noise between {\it logical} qubits performing a given algorithm with $k$ levels of concatenated error correction, 
Appendix~\ref{app:long_range} shows that
\begin{equation}
\label{eq:Deltak-final}
t_0\Delta_{\rm L}^{(k)}=\frac{ \left(2e^{2+1/e}B^2 t_0\Delta^{(0)} \right)^{2^k} }{2e^{2+1/e}B^2} \ D^{\beta 2^k k},
\end{equation}  
where $\Delta^{(0)}$ is the magnitude of the long-range noise between physical qubits performing the same algorithm without error correction (so its logical qubits are its physical qubits).
Eq.~(\ref{eq:Deltak-final}) gives an over-estimate of the true error, but no better bound exists at present. Thus the $k$ that minimizes this bound gives the 
best existing bound on the achievable accuracy in the presence of such noise.
For example, if 
$\Delta^{(0)}$ is the crosstalk between {\it physical} 
qubits in a computer performing a given calculation 
without error correction, then 
$\Delta_{\rm L}^{(k)}$ is an upper-bound on the crosstalk between {\it logical} qubits in a computer performing the same calculation
with $k$ levels of error-correction.
Since Eq.~(\ref{eq:Deltak-final}) has the same $k$-dependence as in Sec.~\ref{subsec:ResConstr},  all results there
and in Appendix~\ref{app:p_kmax} hold for this long-range noise, 
upon replacing $B$ by $2e^{2+1/e}B^2$.
A quick estimation of $D$ is that it equals the number of fault locations in a ``Rec'', so $D\sim A'=291$; a more precise calculation \cite{our-fullstack} confirms that $D$ is of order $A'$.
 
The good news is that this shows that error correction can reduce the errors to a certain extent, even for noise that is too long-ranged to have a fault-tolerance threshold.
The bad news is that one requires $t_0\Delta^{(0)} < \left[ 2e^{2+1/e}B^2D^{2\beta}\right]^{-1}$ for error correction to be useful in reducing crosstalk between qubits. This is always tiny; it is of order $10^{-9}$ for small $\beta$, and of order $10^{-13}$ for 
$\beta=1$. If one achieves noise as weak as this, one can already do huge quantum calculations without worrying about errors.

In this section, we treated scale-dependent crosstalk induced by resource constraints, but our conclusions apply to long-range correlated noise of any origin. One example is unwanted long-range interactions which decays with the distance $r$ between qubits  $i$ and $j$, so  $H_{ij}\propto (1/r)^{z}$. Ref.~\cite{Aharonov2006Feb} considered this example in a $d$-dimensional lattice of qubits for $z>d$, but we treat longer-ranged noise ($z\leq d$) in Appendix~\ref{app:long_range}, for which $\Delta\propto N^{(1-z/d)}$. Then this example is the same as those treated above with $\beta=(1-z/d)$.

\subsection{Energy constraint for resonant gates performing Shor's algorithm}\label{sec:Example}
\label{sec:waveguide}

As a concrete example of the situation described in Sec.~\ref{subsec:ResConstr} (with $\beta=1$), we examine a resource constraint in a specific type of quantum gate implementation, performing a specific quantum algorithm.  
We take the gate implementation to be resonant qubit gates , and 
assume there are limited energy-resources available to perform a given computation (see also a related early analysis in Ref.~\cite{gea2002some}).
We take the desired computation to be the Shor algorithm, and investigate how big a computation can be performed (with a given calculational accuracy) when there are limited energy-resources.

\begin{figure}
\includegraphics[width=0.93\columnwidth]{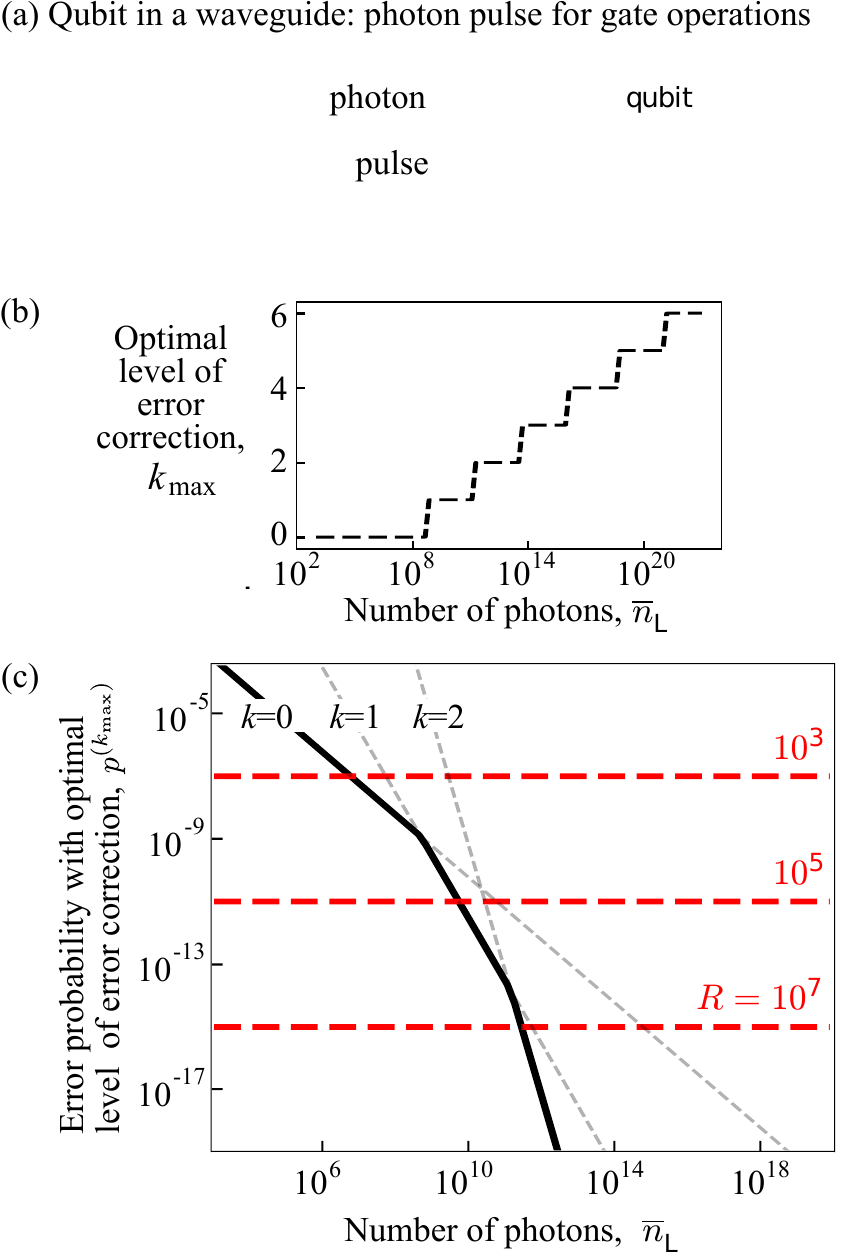}
\caption{\label{fig:Example} (a) Qubit-in-waveguide schematic. (b) $k_{max}$ as a function of number of photons per logical gate $\nbarl=\nbart/R^2$ (see text). (c) Lowest possible error per logical gate $p^{(k_{max})}$ as a function of $\nbarl=\nbart/R^2$, for the Shor's factoring algorithm. The horizontal red dashed lines correspond to the different target ($p_\mathrm{err}$) values for different $R$ values.
}
\end{figure}

We consider qubits embedded in waveguides, i.e., a continuum of electromagnetic modes prepared at zero temperature. Gates are activated by resonant propagating light pulses with a well-defined average energy, or, equivalently, an average photon number $\overline{n}_g$ [see Fig.~\ref{fig:Example}(a)]. This describes the situation in superconducting circuits \cite{krantz2019quantum, peropadre2013scattering} and integrated photonics \cite{lodahl2015interfacing}. It is also the paradigm of quantum networks and light-matter interfaces, with successful implementations in atomic qubits \cite{Review-Rempe}. Here, we maximize the accuracy of a computation for a given energy constraint.
Then, by inverting the logic, we use this to find the minimal energy budget necessary to realize a specific computation with desire accuracy.
For our illustrative goals, we treat only single-qubit gates subjected to noise from spontaneous emission. 
In doing so, we neglect the dephasing noise. While this is a fair approximation for atomic qubits, it is more demanding for solid-state qubits, but within eventual reach of superconducting circuits and spin qubits.

The qubit's dynamics follow the Lindblad equation, $\dot{\rho}=- \frac{\upi}{\hbar} [H(t),\rho]+ \mathcal{D}(\rho)$, with the total Hamiltonian $H(t)=H_0 + H_\mathrm{D}(t)$. Here, $H_0\equiv -\frac{1}{2}\hbar\omega_0\sigma_z$ is the qubit's bare Hamiltonian, with $\sigma_z\equiv |0\rangle\langle 0|-|1\rangle\langle 1|$. 
We assume that the gates are designed as resonant driving in the rotating wave approximation (RWA), the Hamiltonian term for this drive is $H_\mathrm{D}(t)\equiv \frac{\hbar}{2} \Omega h(t) {\left(|0\rangle\langle 1|\upe^{\upi\omega_0 t}+|1\rangle\langle 0|\upe^{-\upi\omega_0 t}\right)}$, with the Rabi frequency $\Omega \ll \omega_0$.
We take $h(t)$ as a square function, nonzero only for the duration of the gate: $h(t)=1$ for $t\in[0, \tau]$ and $0$ otherwise.  The unitary evolution induced by this Hamiltonian is a rotation around the x-axis of the Bloch sphere by an angle $\theta = \Omega \tau$. This rotation is our single-qubit gate.
In addition, the Lindblad dissipator $\mathcal{D}(\rho)\equiv \gamma {\left(\sigma_- \rho \sigma_+ -\frac{1}{2} \{ \rho, \sigma_+ \sigma_- \} \right)}$ accounts for the spontaneous emission in the waveguide. Here $\sigma_-\equiv |0\rangle\langle 1|$ is the lowering operator, $\sigma_+=\sigma_-^\dagger=|1\rangle\langle 0|$ the raising operator, and $\gamma$ the spontaneous emission rate. 

Note that $\Omega$ and $\gamma$ are not independent, typical of waveguide Quantum Electrodynamics, where the driving and the relaxation take place through the same one-dimensional electromagnetic channel. 
Spontaneous emission events while the driving Hamiltonian is turned on cause errors in the gate implementation. Their impact is reduced if the qubit is driven faster, i.e., if the Rabi frequency is larger. Conversely, the Rabi frequency is related to the mean number of photons inside the driving pulse through $\Omega = \frac{4 \gamma}{\theta} \overline{n}_g$ (see Appendix~\ref{app:gates}). In principle, pulses containing more photons induce better gates, with perfect gates for an infinite number of photons.
However, if one designs  the gates to work within RWA (to avoid the complicated pulse-shaping issues that come with finite counter-rotating terms) then
there cannot be too many photon; $\overline{n}_g \ll \omega_0/\gamma$.
Then the remnant noise, for a $\theta=\pi$ gate, has physical error probability (see Appendix~\ref{app:gates})
$\eta = \frac{\pi^2}{16}\frac{1}{\overline{n}_g}$, with a minimal noise of order $\gamma/\omega_0$.

We now assume a constraint on the total number of photons $\nbart$ available to run the whole computation. As we show below, taking this constraint into account allows us to minimize the resource needed, for a target level of tolerable computational error. At level-$k$ concatenation, the number of physical gates needed to implement a computation with $L$ (logical) gates is $LG^{(k)}=LA^k$. Assuming a distinct pulse for each gate, the number of photons available per physical gate, given the total energetic constraint, is $\nbarg=\nbart(LA^k)^{-1}$. Thus, the physical error probability for the $\theta=\pi$ gate acquires an exponential $k$ dependence,
\begin{equation}
    \eta^{(k)}= \frac{\pi^2}{16}\frac{LA^k}{\nbart}.
    \label{eq:JCp}
\end{equation}
Thus this is a concrete example corresponding to the $\beta=1$ case in Eq.~(\ref{Eq:expon}) above.

To better grasp the consequences of this $k$-dependent $\eta$, which we take as the generic behavior for all gates, we consider carrying out Shor's factoring algorithm \cite{Shor94}. Shor's factoring algorithm is touted as the reason the RSA public-key encryption system will be insecure when large-scale  quantum computers become available. The current RSA key length is $R=2048$ bits. The exponential speedup of Shor's algorithm over known classical methods comes from the fact that we can do the discrete Fourier transform on an $R$-bit string using $O(R^2)$ gates on a quantum computer (see, for example, Ref.~\cite{NielsenChuang2010}), compared to $O(R2^R)$ gates on a classical computer. The discrete Fourier transform gives a period-finding routine within the factoring algorithm, the only step that cannot be done efficiently classically. Thus, to run Shor's algorithm, one needs $L\sim R^2$ non-identity logical quantum gates for the discrete Fourier transform. The exact number of computational gates, including the identity gate operations which can be noisy, for the full Shor's algorithm depends on the chosen circuit design and architecture. We will take the lower limit of $L\sim R^2$ in what follows. The concatenation values we find below are thus likely optimistic estimates.

A standard strategy is to demand that the computation runs correctly with probability $P_\mathrm{target}>1/2$; once this is true, the computation can be repeated to exponentially increase the success probability towards 1. For $P_\mathrm{target}=2/3$, with $L$ logical gates, each with error probability $\perr$ ($\ll 1$), we require $(1-\perr)^L>P_\mathrm{target}=2/3$, giving a target error probability per logical gate of $\perr\lesssim(3L)^{-1}$. 

Fig.~\ref{fig:Example}(b) presents $k_{\max}$, the maximum concatenation level, as an increasing function of the photon budget per logical gate $\nbarl=\nbart/R^2 (\sim \nbart/L)$ (note: $\nbarl=G^{(k)}\nbarg=A^k\nbarg$). For fixed total resource, in this case, the photon budget per logical gate, as the concatenation level increases, the available photon count per physical component falls, and we recover the behavior observed in earlier sections, giving a finite $k_{\max}$, and consequently, a limit to the computational accuracy. The solid bold black line in Fig.~\ref{fig:Example}(c) gives the corresponding minimum attainable error per logical gate as a function of $\nbarl$.

\section{Minimizing the resource costs of an algorithm}
\label{sec:min-energy}
One can turn our results for resource constraints around,
to enable us to answer the following question: What are the minimum resources needed for a target computational accuracy, sufficient for given problem?  

As an illustration, we answer this question for the Shor's algorithm for an $R$-bit string in Sec.~\ref{sec:Example}. The number of gates in the algorithm grows with $R$, requiring a smaller $\perr$ for the algorithm to be successful. This demands a larger photon budget to implement the algorithm using resonant gates. For the  parameters in Fig.~\ref{fig:Example}(c), $R = 10^{3}$ requires no concatenation, and the minimum photon budget is $\nbarl=10^6$; for $R = 10^{5}$, we need $k=1$ and $\nbarl=10^{9}$; for $R=10^7$, we need $k=2$ and $\nbarl=10^{11}$. Recall that the gates are assumed to be designed within the RWA, hence $\omega_0\gg \gamma\nbarg$. For $R=10^3$, this translates to the condition $\omega_0\gg \gamma\nbarg =\gamma\nbarl/A^0=10^7\gamma$; for $R=10^7$, we need $\omega_0\gg 10^6\gamma$ for $A=575$ (this $A$ is from Ref.~\cite{Aliferis2006}).  These conditions are attained for atomic qubits. They are within reach of future generations of superconducting qubits, where $\gamma \sim 10$ Hz for qubit frequency $\omega_0 \sim 10$ GHz. Today, the best coherence time for superconducting qubit is within the millisecond range: $\gamma \sim 1$kHz \cite{kjaergaard2019superconducting,sears2013extending}.

Our analysis also provides an estimate of the energy needed to run the gates involved in the Shor's algorithm, namely, $E_\mathrm{tot} \sim \hbar \omega_0 L \nbarl(R)$. For $R=10^3$, with the above photon budget of $10^6$, this translates into $E_\mathrm{tot}\sim$ 1 pJ. Taking into account the parallelization of the computation 
(see Appendix~\ref{app:energeticBill}), this corresponds to a typical power consumption of about 1 pW, while the $R=10^7$ case requires only $10$ nW of power. 

Thus for a realistic constraint on the photon power to perform
quantum gate operations, quantum error correction would allow one to perform large quantum computations.
This is a surprising and positive conclusion, when one considers that the constraint causes scale-dependent errors for which there is no fault-tolerant threshold.  It clearly shows that the absence of a threshold is not necessarily a significant impediment to using error correction in quantum computing.

Here, we have calculated
the energy of photons arriving at the qubits for a gate operation. However, that energy is a fraction of the energy for signal generation because there is an attenuator between the signal generator and the qubit. This attenuator's job is to absorb thermal photons to avoid them perturbing the qubit, however this means it also absorbs most of the photons in the signal sent to perform the gate operation.
To calculate the signal generation energy from the above results, one must multiply by the attenuation.
Unfortunately, the attenuation depends on design choices beyond the discussion here (like the temperature at which the signal generation occurs).  Furthermore, there is a large cryogenic energy budget for keeping the qubits and attenuators cool, which depends on the photon energy absorbed by the attenuator. Elsewhere \cite{our-fullstack} we will perform a full energetic optimization of all of these interlinked components (along with other critical components of a quantum computer) in a {\it full-stack} analysis of a large-scale quantum computer.

Nevertheless, the above example does show the fundamental ingredient in the minimization of energy (or any other resource) is the calculation of the scale-dependence of the noise that occurs when that resource is constrained.  To go from this to a full-stack analysis is mostly an issue of optimizing cryogenics, control circuitry (including signal generation), and the quantum algorithm for the calculation in question.

\section{CONCLUSIONS}\label{sec:conc}

\noindent Many quantum computing technologies currently exhibit physical gate errors that grow with the size and complexity of the quantum computer. 
Then there is no fault-tolerance threshold. 
Despite this, we show that a certain amount of error correction can increase calculational accuracy, but that this accuracy decreases again with too much error correction.  We show how to find the amount of error correction that optimizes this accuracy.
For concreteness, we considered  concatenated 7-qubit codes here.  
However, our approach could be applied to other fault-tolerance schemes (surface codes, measurement-based, etc) \footnote{For such schemes, $k$ will be replaced by the scale of the error correction (i.e. a measure of the number of physical components in the error correction). Then one needs to replace our Eq.~(\ref{eq:pk}) with a relation (or numerical estimate) for the scaling of the logical error strength with $k$, which is not yet available for some schemes.}, where we also expect that scale-dependence of noise on physical components can lead to situations where a little error correction is good, but too much is bad.

We explored the optimization of calcuational accuracy in increasing levels of practical relevance, from a simple toy example, to physical qubits in waveguides. 
We identified some cases, such as reasonable energy constraints for gate operations, where optimization gives a maximum accuracy good enough for large quantum algorithms. In other cases, such as volume or bandwidth constraints causing long-range crosstalk between qubits, error correction is only useful against such crosstalk when the error strength per physical gate is already so small (ranging from $10^{-9}$ to $10^{-13}$) that one could perform huge quantum calculations without any error correction.

Our analysis suggests three priorities for experimenters working towards useful quantum computers: (1) they should try to characterize the scale-dependence of the errors for their technology;
(2) they should strive to make this scale dependence as weak as possible; 
(3) they should reduce the physical error probability significantly below the standard threshold. 
Point (3) is good to make standard fault tolerance work well, but it becomes critical when errors are scale-dependent. 
In this context, the optimization in this work
will enable them to see the size of quantum computation that can be treated with their error magnitude and scale-dependence.  Experimenters will be aided in addressing these points by a full-stack model of a quantum computer \cite{our-fullstack}, built from the theory presented here.

\acknowledgments
This work is supported by a Merlion Project (grant no.~7.06.17), and by the Agence Nationale de la Recherche under the programme ``Investissements d'avenir" (ANR-15-IDEX-02) and the Labex LANEF. HKN also acknowledges support by a Centre for Quantum Technologies (CQT) Fellowship. CQT is a Research Centre of Excellence funded by the Ministry of Education and the National Research Foundation of Singapore.

%
%
%
%
%
%

\appendix

\section{Bounds on $p^{(k_{\rm max})}$}
\label{app:p_kmax}

\begin{figure*}
\includegraphics[width=\textwidth]{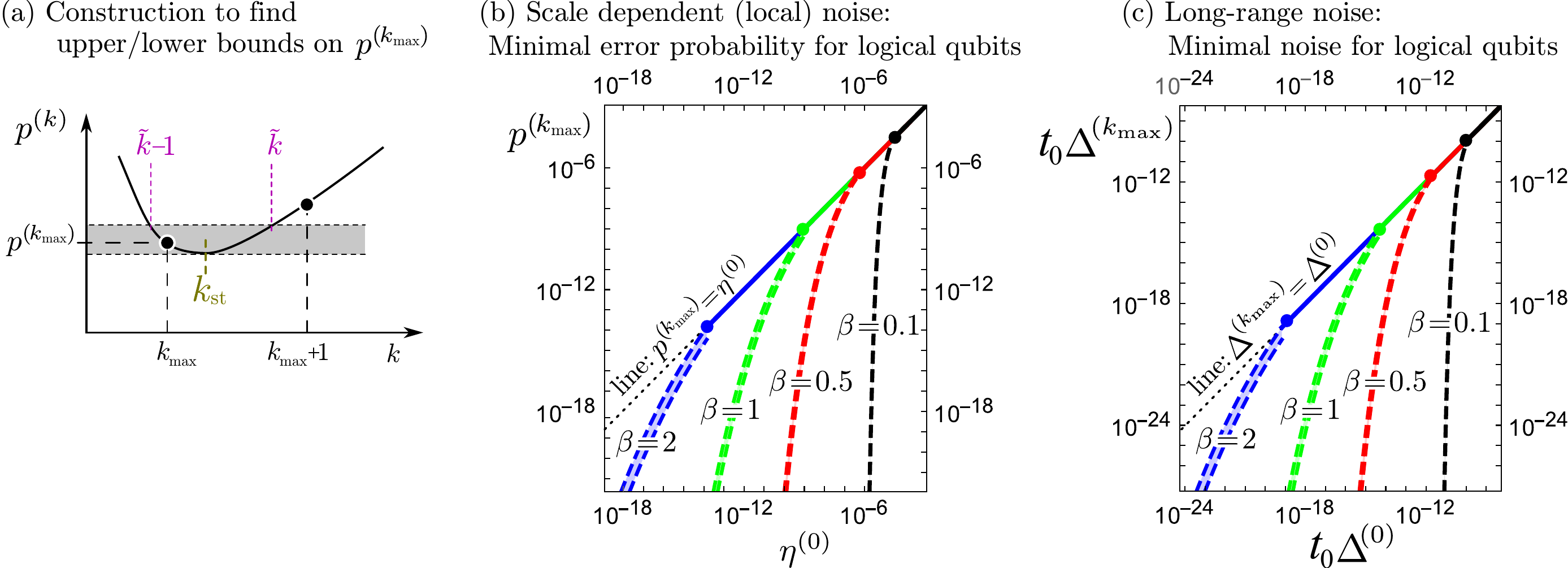}
\caption{\label{fig:p_kmax}
(a) Sketch of  $p^{(k)}$ (black curve), with integer $k$ marked by black filled-circles.
Then $p^{(k_{\rm max})}$ must be between $p^{(k_{\rm st})}$ and $p^{(\tilde k)}$ (the gray region).
(b) A plot of $p^{(k_{\rm max})}$ with $D=291$ and $B=10^4$, for  $\beta=0.1$ (black), $\beta=0.5$ (red), $\beta=1$ (green) 
and $\beta=2$ (blue).  
The filled-circles mark  $\eta^{(0)}= B^{-1}D^{-2\beta}$ for each $\beta$.
For  $\eta^{(0)}\geq B^{-1}D^{-2\beta}$, concatenation is useless,
and $p^{(k_{\rm max})}=\eta^{(0)}$ (solid lines).
For $\eta^{(0)}<B^{-1}D^{-2\beta}$, concatenation is useful, 
and $p^{(k_{\rm max})}$  is in the shaded region 
between the two dashed curves of the same color, the upper being $p^{(\tilde k)}$ and 
the lower being $p^{(k_{\rm st})}$.
The two curves are close enough to give a good estimate  of $p^{(k_{\rm max})}$;
indeed for $\beta<0.5$ the two curves are almost indistinguishable.
(c) The same as (b) but for the long-range noise given by Eq.~(\ref{eq:Deltak}). The curves look similar to (b) but the magnitudes on the axes are much smaller.
} 
\end{figure*}

Here we give algebraic results for $k_{\rm max}$ and $p^{(k_{\rm max})}$, 
when the scale-dependent noise is given by Eq.~(7) in the main article.
For $\eta^{(0)} \geq B^{-1}D^{-2\beta}$, the noise is too strong for concatenation to be useful at all.  Then $k_{\rm max}=0$, meaning logical gates are physical gates, so the minimal error probability of a logical qubit $p^{(k_{\rm max})} = \eta_0$.
For $\eta^{(0)} > B^{-1}D^{-2\beta}$, concatenation is useful to reduce errors, and $k_{\rm max} \geq 1$.
Then there is no simple algebraic form for $p^{(k_{\rm max})}$.  However, we can find algebraic formulas for fairly close upper/lower bounds on $p^{(k_{\rm max})}$, using the fact that Eq.~(7) in the main article gives a curve for $p^{(k)}$ for all $k$, 
even if only integer $k$ are physically meaningful [see Fig.~\ref{fig:p_kmax}(a) below].
The minimum of $p^{(k)}$ will typically be at a non-integer value of $k$, so
defining this minimum as $k_{\rm st}$ (``st'' because it is a stationary point),
we have 
\begin{eqnarray}
k_{\rm st} = -\frac{1}{\ln[2]}  - \frac{\ln\big[B\eta^{(0)}]}{\beta\ln[D]} \,,
\end{eqnarray}
which means that
\begin{eqnarray}
p^{(k_{\rm max})}\geq p^{(k_{\rm st})}=
\frac{1}{B} \exp\left[- \frac{\beta}{g_2}  \exp\left[-1- \frac{g_2 \ln\big[B\eta^{(0)}]}{\beta}  \right]\right]
\nonumber \\
\label{Eq:p_kmax_lower}
\end{eqnarray}
where $g_2=\ln(2)/\ln (D)$.

To find an upper bound on $p^{(k_{\rm max})}$, we note that $p^{(k)}$ is a convex function of $k$ (i.e. $d^2p^{(k)}/dk^2 \geq 0$ for all $k\geq0$), 
so we can uniquely define $\tilde k$ such that $p^{(\tilde k)}=p^{(\tilde k-1)}$,
so
\begin{eqnarray}
\tilde k = - \ln {\left(B\eta^{(0)}D^\beta\right)}\big/ \ln{\left(D^\beta\right)}.
\end{eqnarray}
We then know that $(\tilde k-1) \leq k_{\rm max} \leq \tilde k$, see the sketch in
Fig.~\ref{fig:p_kmax}(a) below, thereby giving Eq.~(8) in the main article.
In addition,  we have
\begin{equation}
p^{(k_{\max})}  \leq p^{(\tilde k)} =
\frac{1}{B} \exp\left[- g_1\beta  \left(B\eta^{(0)}\right)^{-g_2/\beta}\right],
\label{Eq:p_kmax_upper}
\end{equation}
where $g_1=\ln[D]/2$. 

The upper and lower bounds on $p^{(k_{\rm max})}$  in Eq.~(\ref{Eq:p_kmax_upper})
and Eq.~(\ref{Eq:p_kmax_lower}) are shown in Fig.~\ref{fig:p_kmax}(b) below.
This shows that the bounds are close enough that either formula gives
a good estimate of the minimal probability of an error in a logical qubit, 
$p^{(k_{\rm max})}$ for realistic $D, B, \eta^{(0)}$, and $\beta$.

\section{Details for long-range correlated noise}\label{app:long_range}

Here we detail our analysis of long-range correlated noise, which uses the method of Refs.~\cite{Aharonov2006Feb,Preskill2013Mar}, but applies it to cases where Refs.~\cite{Aharonov2006Feb,Preskill2013Mar} showed that the noise was too long-ranged to get a fault tolerance threshold.
Naively, one might think this means error correction is useless in these cases, but this is not so.

Ref.~\cite{Aharonov2006Feb} proved that quantum error correction will correct errors due
to the long-range noise in Eq.~(\ref{Eq:Aharonov-Kitaev-Preskill}), 
at least as well as it will correct local-noise of strength $e^{1+1/(2e)}\sqrt{2t_0\Delta}$.
Hence, one can take Eq.~(\ref{eq:pk}) and replace $\eta$  with  $e^{1+1/(2e)}\sqrt{2t_0\Delta}$,
and replace $p^{(k)}$ with $e^{1+1/(2e)}\sqrt{2t_0\Delta^{(k)}}$, where we define $\Delta^{(k)}$ as the effective long-range noise between the logical qubits after $k$-level of concatenation.
So the upper bound on the long-range noise between {\it logical} qubits after k-levels of concatenation is
\begin{equation}
\label{eq:Deltak}
t_0\Delta_{\rm L}^{(k)}=\frac{ \left(2e^{2+1/e}B^2 t_0\Delta \right)^{2^k} }{2e^{2+1/e}B^2}.
\end{equation}
where we note that $2e^{2+1/e}\sim 21$.
Refs.~\cite{Aharonov2006Feb,Preskill2013Mar} considered $\Delta$ to be finite and independent of the number of physical qubits, $N$, in the limit of large $N$. Then Eq.~(\ref{eq:Deltak}) gives a fault-tolerance threshold at $t_0\Delta = (2e^{1+2/e}B^2)^{-1}\sim 10^{-9}$.
In contrast, we consider $\Delta$ that grows with $N$, and so grows with $k$ like 
$N \simeq D^k N_0$ where $N_0$ is the number of qubits required to perform the algorithm without any error correction.   Here $D$ is defined by saying that each additional level of concatenation replaces each logical qubit with $D$ logical qubits.   A very rough estimate of $D$ shows that it is of order the number of gates in a ``Rec'' so $D\sim A'=291$.  
A more detailed calculation of $D$ confirms that it is of the same order of magnitude as $A'$ \cite{our-fullstack}.

In principle, one could imagine an arbitrary dependence of $\Delta$ on $N$, and hence on $k$. Then the optimal amount of error correction for such long-range noise would be that given above in Sec.~\ref{subsec:Gen}, with $B$ replaced by $2e^{2+1/e}B^2$.
If  $\Delta \propto N^\beta$ one substitutes $\Delta= \Delta^{(0)} D^{\beta k}$ into the right hand side of Eq.~(\ref{eq:Deltak}),
where $\Delta^{(0)}$ is the magnitude of the long-range noise when there is no error correction, for which there are only $N_0$ physical qubits.
This gives Eq.~\eqref{eq:Deltak-final},
which has the same $k$-dependence as in Sec.~\ref{subsec:ResConstr}. Hence, all results in Sec.~\ref{subsec:ResConstr}
and Appendix~\ref{app:p_kmax} hold for long-range noise, 
so long as one replaces $B$ by $2e^{2+1/e}B^2$.

We then use the results in  Appendix~\ref{app:p_kmax} to plot the minimal value of $t_0\Delta_{\rm L}^{(k)}$ in  Fig.~\ref{fig:p_kmax}c. 
While we took the rough estimate of $D$ give above ($D=291$) for the plots, we observed that the form of the curves in 
 Fig.~\ref{fig:p_kmax}b,c was rather insensitive to the exact value of $D$.

We now turn to an example given in Ref.~\cite{Aharonov2006Feb}, in which it was assume the qubits were placed on a $d$-dimensional lattice, with the unwanted interaction between qubits at position $r_i$ and $r_j$ being
\begin{eqnarray}
||H_{ij}|| = \delta\,|r_i-r_j|^{-z}.
\label{Eq:H_decay_with_distance} 
\end{eqnarray} 
Ref.~\cite{Aharonov2006Feb} considered this model when the noise was not too long-ranged ($z>d$) so that $\Delta$ remains finite as $N\to \infty$.
However, in many designs of quantum computer, one has circuit elements that perform two qubit gates between physically distance qubits.  Noise in such circuit elements could generate even longer-range noise than considered in Ref.~\cite{Aharonov2006Feb}.  Such noise may not always be a simple function of $(r_i-r_j)$, but we can get a feel for such very long range noise by taking Eq.~(\ref{Eq:H_decay_with_distance}) with $z\leq d$.
If we assume that $N\gg 1$ and that the nature of the lattice (i.e., its dimensionality, aspect ratio, etc) is unchanged as we increase $N$, then 
$\Delta \propto N^{1-z/d}$ for $z<d$.
This then gives Eq.~\eqref{eq:Deltak-final} with $\beta=(1-z/d)$.  To calculate $\Delta_0$ from $\delta$, one must take a concrete example, such as a chain of $N_0$ qubits in one dimension, or a $\sqrt{N_0}\times\sqrt{N_0}$ square lattice in two dimensions. Then $\Delta_0$ is given by the central qubit in the lattice (since it has the largest unwanted coupling to other qubits), and in the limit $N_0 \gg 1$ one gets the results in table~\ref{tab:scale_dep_noise}.   For $z>d$ this has the scaling discussed in Sec.~\ref{sec:long-range} with $\beta=(1-z/d)$. In the special case of $z=d$, one has a $k$-dependence that coincided with the toy-model in Sec.~\ref{subsec:toy}.

\begin{table*}
\begin{tabular}{|l|c|c|c|c|}
\hline
 & & \multicolumn{2}{|c|}{Parameters for $d$ dimensional array of qubits}\\
 & Scaling with $k$  & \multicolumn{1}{|c}{\ \ $d=1$ (chain)\ \ } & \multicolumn{1}{c|}{\ \ $d=2$ (square lattice)\ \ } 
\\
\hline
\ $z < d$\ \ 
& Eq.~(\ref{eq:Deltak-final}) with $\beta=1-z/d$
& \ \ \ \phantom{${\Big|\atop \Big|}$}
${\displaystyle \Delta_0=\frac{\delta}{a^z}\,\frac{2^{z}N_0^{1-z}}{1-z} }$ \ \ \ \ 
&\ \ \ ${\displaystyle \Delta_0=\frac{\delta}{a^z} \frac{2^{z+1}N_0^{1-z/2}C_z}{2-z}} $ \ \ \
\\
\hline
\ $z=d$\ \ 
& \ \ \ ${\displaystyle \phantom{{\Bigg|\atop \Big|}} \Delta^{(k)}t_0= \frac{ \left(2e^{2+1/e}B^2\, t_0\delta\big/a^d \right)^{2^k}}{2e^{2+1/e}B^2}  \big(C_0  + \ln[D] k\big)^{2^k}} $ \ \ \ 
& \phantom{${\Big|\atop \Big|}$} ${\displaystyle C_0=2 \ln\left[\kappa_1\frac{N_0}{2}\right]}$  
& ${\displaystyle C_0=\pi \ln\left[\kappa_2\frac{N_0}{4}\right]}$\\
\hline
\end{tabular}
\caption{\label{tab:scale_dep_noise} Parameter dependence for the long-range interaction model with $z \leq d$ in Eq.~(\ref{Eq:H_decay_with_distance}). We assume large $N_0$, so the sum over $j$ in $\Delta$ can be approximated by an integral.  The $z$-dependent constant for the square lattice $C_z=\int_0^{\pi/4} \cos^{z-2}\theta$. For $z=d$, there are order-one constants, $\kappa_1$ and $\kappa_2$, which can be neglected for large $N_0$. These constants come from the integral's short distance cutting-off, and a precise calculation of them would require not approximating the sum as an integral.}
\end{table*}

\section{Resonant qubit gates}\label{app:gates}
We consider a two-level system---the qubit---with bare Hamiltonian $H_0=-\frac{1}{2}\hbar \omega_0 \sigma_z$ embedded into a waveguide for light at resonant frequency $\omega_0$ for implementing gate operations on the qubit. Assuming the system is at $0K$, and neglecting pure dephasing, the physics is described by the optical Bloch equations in which there is only spontaneous emission. The driving Hamiltonian is $H_\mathrm{D}(t)\equiv \hbar \Omega h(t)\cos(\omega_0 t) \sigma_x$, writable in the form given in the main text, $H_\mathrm{D}(t)\rightarrow \frac{1}{2}\hbar\Omega(t)(|0\rangle\langle 1|\upe^{\upi\omega_0 t}+|1\rangle\langle 0|\upe^{-\upi\omega_0 t}$ under the RWA. The overall qubit dynamics follows the Lindblad equation given in the main text: $\dot\rho =-\frac{\upi}{\hbar}[H(t),\rho] + \mathcal{D}(\rho)$, with $H(t)\equiv H_0+H_\mathrm{D}(t)$ and $\mathcal{D}(\rho)$ the dissipator defined as $\mathcal{D}(\rho)=\gamma(\sigma_-\rho\sigma_+-\frac{1}{2}\{\rho,\sigma_+\sigma_-\})$.

The gate on the qubit is accomplished by an incoming coherent light pulse of power $P_{in}=\hbar \omega_0 \dot{N}_\mathrm{in}$ where $\dot{N}_\mathrm{in}$ is the rate of incoming photons. The Rabi frequency induced by pulse is $\Omega=2 (\gamma \dot{N}_{in})^{1/2}$ \cite{cottet2017observing}. It increases with $\gamma$, the time-constant for spontaneous emission,  as both quantities measure the strength of the coupling between the qubit and the modes of the waveguide which provide both the decay and driving channels. As we are considering energetic constraints, i.e., a limit on the total number of photons to do gates, it is useful to express $\Omega$ in terms of the  photon number $\nbarg$ available for that gate. For $H_D(t)$ describing a square pulse of duration $\tau$ with constant power, with $\nbarg$ available photons, $\dot{N}_\mathrm{in}=\nbarg/\tau$. In addition, to induce a rotation angle of $\theta$, we require $\Omega \tau =\theta$, so that $\tau=\theta/\Omega$. We thus have $\Omega=4 \gamma \nbarg/\theta$,
and $\tau=\theta^2/(4\gamma\nbarg)$ when expressed in terms of given $\nbarg$ and $\theta$. Observe that, for a target $\theta$, larger input energy, i.e., larger $\nbarg$, enables faster gate operation.

The Lindblad equation, together with the expressions for $\Omega$ and $\tau$ in terms of $\theta$ and $\nbarg$, describes the noisy implementation of a rotation of the qubit state by angle $\theta$ about the $x$ axis in the Bloch ball, using given energy $\hbar\omega_0\nbarg$. The noisy gate operation, $\widetilde{\mathcal{G}}$, obtained by integrating the Lindblad equation over the gate duration $\tau$, is a linear map that takes the input qubit state $\rho(t=0)$ to the (noisy) output state $\rho(t=\tau)$. It can be written in terms of the ideal gate $\mathcal{G}$ as $\widetilde{\mathcal{G}}=\mathcal{G}\circ\mathcal{E}$, with the noise map $\mathcal{E}\equiv \mathcal{G}^{-1}\circ\widetilde{\mathcal{G}}$. $\mathcal{E}$ is a completely positive (CP) and trace-preserving (TP) linear map, writable, using the Pauli operator basis $\{\id, \sigma_x,\sigma_y,\sigma_z\}=\{\sigma_\alpha\}_{\alpha=0}^3$ (with $\sigma_0=\id,\sigma_1=\sigma_x$, etc.), as
\begin{align}
\mathcal{E}(\rho)=\sum_{\alpha,\beta=0}^3 \chi_{\alpha\beta}\, \sigma_\alpha \rho \sigma_\beta,
\end{align}
where $\chi_{\alpha\beta}$ are scalar coefficients.

The coefficients $\chi_{11}\equiv p_x, \chi_{22}\equiv p_y$, and $\chi_{33}\equiv p_z$ give the probabilities of $X$, $Y$, and $Z$ errors, respectively, relevant for the 7-qubit code used in our discussion ($\chi_{\alpha\beta}$, with $\alpha\neq \beta$, do not affect the code performance; see, for example, Ref.~\cite{NielsenChuang2010}). Straightforward calculation gives $ \frac{1}{2}\tr{\left\{\sigma_\alpha\mathcal{E}(\sigma_\alpha)\right\}}=\chi_{00}+\chi_{\alpha\alpha}-\sum_{\beta\neq 0,\alpha}\chi_{\beta\beta}$, for $\alpha=0,1,2,3$. We obtain $p_x$, $p_y$, and $p_z$ by inverting these relations. For $\theta=\pi$, corresponding to the commonly used gate $\mathcal{G}=X(\cdot)X$, we find
\begin{equation}
p_x\simeq \frac{\pi^2}{16} \frac{1}{\nbarg}, \quad p_y  \simeq \frac{\pi^2}{32} \frac{1}{\nbarg}, \quad\textrm{and}\quad p_z \simeq  \frac{\pi^2}{32} \frac{1}{\nbarg}\,,
\end{equation}
accurate to linear order in $1/\nbarg$.
The largest of these, namely, $p_x$ is what we set as $\eta$ in the main text.

\begin{table}[t]
\caption{\label{tab:energeticBill}Energetic bill for carrying out Shor's algorithm in our qubit-in-waveguide example.}
\begin{tabular}{c|c|c|c}
& $R=10^3$ & $R=10^5$ & $R=10^7$ \\
\hline
\hline
Photon number $\nbarl$ & $10^{6}$ & $10^{9}$ & $10^{11}$\\
\hline
Concatenation level $k$ & 0 & 1 & 2\\
\hline
Energy $E_\mathrm{tot}$ & $1$ pJ & $ 10$ $\mu$J & $10$ J \\
\hline
Power $P$ & $ 1$ pW & $ 1$ nW & $ 10$ nW \\
\hline
Total time $L\tau_\mathrm{L}$ & $ 100$ ms & $ 1000$ s & $ 10^9$ s \\
\hline
Gate time $\tau_\mathrm{g}$ & $ 100$ ns & $ 100$ ns & $ 1$ $\mu$s
\end{tabular}
\end{table}

\section{Energetic bill}
\label{app:energeticBill}

Our analysis gives the energy required to carry out Shor's algorithm in a qubit-in-waveguide implementation, for a given problem size $R$: $E_\mathrm{tot}=\hbar\omega_0 L \nbarl(R)\sim\hbar\omega_0R^2\nbarl(R)$, where, as in the main text, $\nbarl(R)$ is the number of photons required per logical gate operation for given $R$ (and hence a given target logical error probability $p_\mathrm{err}$; see main text). $\nbarl(R)$ 
can be read off Fig.~5(c) in the main article.

We can also estimate the average power cost, by assuming that all the logical gates in Shor's algorithm are run sequentially. Each logical gate is assumed to take $M$ clock cycles per concatenation level; $M=3$ for the scheme of Ref. \cite{Aliferis2006}. The duration of a logical gate with $k$ levels of concatenation is thus $\tau_\mathrm{L}=M^k\tau_\mathrm{g}$, with $\tau_\mathrm{g} = \pi^2/(4\gamma \nbarg)$ as the clock interval, taken to be the duration of the $\pi$-pulse gate analyzed above. The power $P$ associated with the energy $E_\mathrm{tot}$ can hence be estimated as $E_\mathrm{tot}/(L\tau_\mathrm{L})$. Some energetic numbers (orders of magnitude only) for the scheme of Ref.~\cite{Aliferis2006}, with $\gamma=10$ Hz and $\omega_0=10$ GHz, are given in Table~\ref{tab:energeticBill}.


\begin{thebibliography}{46}%
\makeatletter
\providecommand \@ifxundefined [1]{%
 \@ifx{#1\undefined}
}%
\providecommand \@ifnum [1]{%
 \ifnum #1\expandafter \@firstoftwo
 \else \expandafter \@secondoftwo
 \fi
}%
\providecommand \@ifx [1]{%
 \ifx #1\expandafter \@firstoftwo
 \else \expandafter \@secondoftwo
 \fi
}%
\providecommand \natexlab [1]{#1}%
\providecommand \enquote  [1]{``#1''}%
\providecommand \bibnamefont  [1]{#1}%
\providecommand \bibfnamefont [1]{#1}%
\providecommand \citenamefont [1]{#1}%
\providecommand \href@noop [0]{\@secondoftwo}%
\providecommand \href [0]{\begingroup \@sanitize@url \@href}%
\providecommand \@href[1]{\@@startlink{#1}\@@href}%
\providecommand \@@href[1]{\endgroup#1\@@endlink}%
\providecommand \@sanitize@url [0]{\catcode `\\12\catcode `\$12\catcode
  `\&12\catcode `\#12\catcode `\^12\catcode `\_12\catcode `\%12\relax}%
\providecommand \@@startlink[1]{}%
\providecommand \@@endlink[0]{}%
\providecommand \url  [0]{\begingroup\@sanitize@url \@url }%
\providecommand \@url [1]{\endgroup\@href {#1}{\urlprefix }}%
\providecommand \urlprefix  [0]{URL }%
\providecommand \Eprint [0]{\href }%
\providecommand \doibase [0]{http://dx.doi.org/}%
\providecommand \selectlanguage [0]{\@gobble}%
\providecommand \bibinfo  [0]{\@secondoftwo}%
\providecommand \bibfield  [0]{\@secondoftwo}%
\providecommand \translation [1]{[#1]}%
\providecommand \BibitemOpen [0]{}%
\providecommand \bibitemStop [0]{}%
\providecommand \bibitemNoStop [0]{.\EOS\space}%
\providecommand \EOS [0]{\spacefactor3000\relax}%
\providecommand \BibitemShut  [1]{\csname bibitem#1\endcsname}%
\let\auto@bib@innerbib\@empty
\bibitem [{\citenamefont {Arute}\ \emph {et~al.}(2019)\citenamefont {Arute},
  \citenamefont {Arya}, \citenamefont {Babbush}, \citenamefont {Bacon},
  \citenamefont {Bardin}, \citenamefont {Barends}, \citenamefont {Biswas},
  \citenamefont {Boixo}, \citenamefont {Brandao}, \citenamefont {Buell},
  \citenamefont {Burkett}, \citenamefont {Chen}, \citenamefont {Chen},
  \citenamefont {Chiaro}, \citenamefont {Collins}, \citenamefont {Courtney},
  \citenamefont {Dunsworth}, \citenamefont {Farhi}, \citenamefont {Foxen},
  \citenamefont {Fowler}, \citenamefont {Gidney}, \citenamefont {Giustina},
  \citenamefont {Graff}, \citenamefont {Guerin}, \citenamefont {Habegger},
  \citenamefont {Harrigan}, \citenamefont {Hartmann}, \citenamefont {Ho},
  \citenamefont {Hoffmann}, \citenamefont {Huang}, \citenamefont {Humble},
  \citenamefont {Isakov}, \citenamefont {Jeffrey}, \citenamefont {Jiang},
  \citenamefont {Kafri}, \citenamefont {Kechedzhi}, \citenamefont {Kelly},
  \citenamefont {Klimov}, \citenamefont {Knysh}, \citenamefont {Korotkov},
  \citenamefont {Kostritsa}, \citenamefont {Landhuis}, \citenamefont
  {Lindmark}, \citenamefont {Lucero}, \citenamefont {Lyakh}, \citenamefont
  {Mandr{\`a}}, \citenamefont {McClean}, \citenamefont {McEwen}, \citenamefont
  {Megrant}, \citenamefont {Mi}, \citenamefont {Michielsen}, \citenamefont
  {Mohseni}, \citenamefont {Mutus}, \citenamefont {Naaman}, \citenamefont
  {Neeley}, \citenamefont {Neill}, \citenamefont {Niu}, \citenamefont {Ostby},
  \citenamefont {Petukhov}, \citenamefont {Platt}, \citenamefont {Quintana},
  \citenamefont {Rieffel}, \citenamefont {Roushan}, \citenamefont {Rubin},
  \citenamefont {Sank}, \citenamefont {Satzinger}, \citenamefont {Smelyanskiy},
  \citenamefont {Sung}, \citenamefont {Trevithick}, \citenamefont
  {Vainsencher}, \citenamefont {Villalonga}, \citenamefont {White},
  \citenamefont {Yao}, \citenamefont {Yeh}, \citenamefont {Zalcman},
  \citenamefont {Neven},\ and\ \citenamefont {Martinis}}]{Google2019}%
  \BibitemOpen
  \bibfield  {author} {\bibinfo {author} {\bibfnamefont {Frank}\ \bibnamefont
  {Arute}}, \bibinfo {author} {\bibfnamefont {Kunal}\ \bibnamefont {Arya}},
  \bibinfo {author} {\bibfnamefont {Ryan}\ \bibnamefont {Babbush}}, \bibinfo
  {author} {\bibfnamefont {Dave}\ \bibnamefont {Bacon}}, \bibinfo {author}
  {\bibfnamefont {Joseph~C.}\ \bibnamefont {Bardin}}, \bibinfo {author}
  {\bibfnamefont {Rami}\ \bibnamefont {Barends}}, \bibinfo {author}
  {\bibfnamefont {Rupak}\ \bibnamefont {Biswas}}, \bibinfo {author}
  {\bibfnamefont {Sergio}\ \bibnamefont {Boixo}}, \bibinfo {author}
  {\bibfnamefont {Fernando G. S.~L.}\ \bibnamefont {Brandao}}, \bibinfo
  {author} {\bibfnamefont {David~A.}\ \bibnamefont {Buell}}, \bibinfo {author}
  {\bibfnamefont {Brian}\ \bibnamefont {Burkett}}, \bibinfo {author}
  {\bibfnamefont {Yu}~\bibnamefont {Chen}}, \bibinfo {author} {\bibfnamefont
  {Zijun}\ \bibnamefont {Chen}}, \bibinfo {author} {\bibfnamefont {Ben}\
  \bibnamefont {Chiaro}}, \bibinfo {author} {\bibfnamefont {Roberto}\
  \bibnamefont {Collins}}, \bibinfo {author} {\bibfnamefont {William}\
  \bibnamefont {Courtney}}, \bibinfo {author} {\bibfnamefont {Andrew}\
  \bibnamefont {Dunsworth}}, \bibinfo {author} {\bibfnamefont {Edward}\
  \bibnamefont {Farhi}}, \bibinfo {author} {\bibfnamefont {Brooks}\
  \bibnamefont {Foxen}}, \bibinfo {author} {\bibfnamefont {Austin}\
  \bibnamefont {Fowler}}, \bibinfo {author} {\bibfnamefont {Craig}\
  \bibnamefont {Gidney}}, \bibinfo {author} {\bibfnamefont {Marissa}\
  \bibnamefont {Giustina}}, \bibinfo {author} {\bibfnamefont {Rob}\
  \bibnamefont {Graff}}, \bibinfo {author} {\bibfnamefont {Keith}\ \bibnamefont
  {Guerin}}, \bibinfo {author} {\bibfnamefont {Steve}\ \bibnamefont
  {Habegger}}, \bibinfo {author} {\bibfnamefont {Matthew~P.}\ \bibnamefont
  {Harrigan}}, \bibinfo {author} {\bibfnamefont {Michael~J.}\ \bibnamefont
  {Hartmann}}, \bibinfo {author} {\bibfnamefont {Alan}\ \bibnamefont {Ho}},
  \bibinfo {author} {\bibfnamefont {Markus}\ \bibnamefont {Hoffmann}}, \bibinfo
  {author} {\bibfnamefont {Trent}\ \bibnamefont {Huang}}, \bibinfo {author}
  {\bibfnamefont {Travis~S.}\ \bibnamefont {Humble}}, \bibinfo {author}
  {\bibfnamefont {Sergei~V.}\ \bibnamefont {Isakov}}, \bibinfo {author}
  {\bibfnamefont {Evan}\ \bibnamefont {Jeffrey}}, \bibinfo {author}
  {\bibfnamefont {Zhang}\ \bibnamefont {Jiang}}, \bibinfo {author}
  {\bibfnamefont {Dvir}\ \bibnamefont {Kafri}}, \bibinfo {author}
  {\bibfnamefont {Kostyantyn}\ \bibnamefont {Kechedzhi}}, \bibinfo {author}
  {\bibfnamefont {Julian}\ \bibnamefont {Kelly}}, \bibinfo {author}
  {\bibfnamefont {Paul~V.}\ \bibnamefont {Klimov}}, \bibinfo {author}
  {\bibfnamefont {Sergey}\ \bibnamefont {Knysh}}, \bibinfo {author}
  {\bibfnamefont {Alexander}\ \bibnamefont {Korotkov}}, \bibinfo {author}
  {\bibfnamefont {Fedor}\ \bibnamefont {Kostritsa}}, \bibinfo {author}
  {\bibfnamefont {David}\ \bibnamefont {Landhuis}}, \bibinfo {author}
  {\bibfnamefont {Mike}\ \bibnamefont {Lindmark}}, \bibinfo {author}
  {\bibfnamefont {Erik}\ \bibnamefont {Lucero}}, \bibinfo {author}
  {\bibfnamefont {Dmitry}\ \bibnamefont {Lyakh}}, \bibinfo {author}
  {\bibfnamefont {Salvatore}\ \bibnamefont {Mandr{\`a}}}, \bibinfo {author}
  {\bibfnamefont {Jarrod~R.}\ \bibnamefont {McClean}}, \bibinfo {author}
  {\bibfnamefont {Matthew}\ \bibnamefont {McEwen}}, \bibinfo {author}
  {\bibfnamefont {Anthony}\ \bibnamefont {Megrant}}, \bibinfo {author}
  {\bibfnamefont {Xiao}\ \bibnamefont {Mi}}, \bibinfo {author} {\bibfnamefont
  {Kristel}\ \bibnamefont {Michielsen}}, \bibinfo {author} {\bibfnamefont
  {Masoud}\ \bibnamefont {Mohseni}}, \bibinfo {author} {\bibfnamefont {Josh}\
  \bibnamefont {Mutus}}, \bibinfo {author} {\bibfnamefont {Ofer}\ \bibnamefont
  {Naaman}}, \bibinfo {author} {\bibfnamefont {Matthew}\ \bibnamefont
  {Neeley}}, \bibinfo {author} {\bibfnamefont {Charles}\ \bibnamefont {Neill}},
  \bibinfo {author} {\bibfnamefont {Murphy~Yuezhen}\ \bibnamefont {Niu}},
  \bibinfo {author} {\bibfnamefont {Eric}\ \bibnamefont {Ostby}}, \bibinfo
  {author} {\bibfnamefont {Andre}\ \bibnamefont {Petukhov}}, \bibinfo {author}
  {\bibfnamefont {John~C.}\ \bibnamefont {Platt}}, \bibinfo {author}
  {\bibfnamefont {Chris}\ \bibnamefont {Quintana}}, \bibinfo {author}
  {\bibfnamefont {Eleanor~G.}\ \bibnamefont {Rieffel}}, \bibinfo {author}
  {\bibfnamefont {Pedram}\ \bibnamefont {Roushan}}, \bibinfo {author}
  {\bibfnamefont {Nicholas~C.}\ \bibnamefont {Rubin}}, \bibinfo {author}
  {\bibfnamefont {Daniel}\ \bibnamefont {Sank}}, \bibinfo {author}
  {\bibfnamefont {Kevin~J.}\ \bibnamefont {Satzinger}}, \bibinfo {author}
  {\bibfnamefont {Vadim}\ \bibnamefont {Smelyanskiy}}, \bibinfo {author}
  {\bibfnamefont {Kevin~J.}\ \bibnamefont {Sung}}, \bibinfo {author}
  {\bibfnamefont {Matthew~D.}\ \bibnamefont {Trevithick}}, \bibinfo {author}
  {\bibfnamefont {Amit}\ \bibnamefont {Vainsencher}}, \bibinfo {author}
  {\bibfnamefont {Benjamin}\ \bibnamefont {Villalonga}}, \bibinfo {author}
  {\bibfnamefont {Theodore}\ \bibnamefont {White}}, \bibinfo {author}
  {\bibfnamefont {Z.~Jamie}\ \bibnamefont {Yao}}, \bibinfo {author}
  {\bibfnamefont {Ping}\ \bibnamefont {Yeh}}, \bibinfo {author} {\bibfnamefont
  {Adam}\ \bibnamefont {Zalcman}}, \bibinfo {author} {\bibfnamefont {Hartmut}\
  \bibnamefont {Neven}}, \ and\ \bibinfo {author} {\bibfnamefont {John~M.}\
  \bibnamefont {Martinis}},\ }\bibfield  {title} {\enquote {\bibinfo {title}
  {Quantum supremacy using a programmable superconducting processor},}\ }\href
  {\doibase 10.1038/s41586-019-1666-5} {\bibfield  {journal} {\bibinfo
  {journal} {Nature}\ }\textbf {\bibinfo {volume} {574}},\ \bibinfo {pages}
  {505--510} (\bibinfo {year} {2019})}\BibitemShut {NoStop}%
\bibitem [{\citenamefont {{Shor}}(1996)}]{Shor1996}%
  \BibitemOpen
  \bibfield  {author} {\bibinfo {author} {\bibfnamefont {P.~W.}\ \bibnamefont
  {{Shor}}},\ }\bibfield  {title} {\enquote {\bibinfo {title} {Fault-tolerant
  quantum computation},}\ }in\ \href {\doibase 10.1109/SFCS.1996.548464} {\emph
  {\bibinfo {booktitle} {Proceedings of 37th Conference on Foundations of
  Computer Science}}}\ (\bibinfo {year} {1996})\ pp.\ \bibinfo {pages}
  {56--65},\ \bibinfo {note}
  {{\href{https://arxiv.org/abs/quant-ph/9605011}{arXiv:quant-ph/9605011}
  }}\BibitemShut {NoStop}%
\bibitem [{\citenamefont {Kitaev}(1997)}]{Kitaev1997}%
  \BibitemOpen
  \bibfield  {author} {\bibinfo {author} {\bibfnamefont {A.~Yu.}\ \bibnamefont
  {Kitaev}},\ }\bibfield  {title} {\enquote {\bibinfo {title} {Quantum
  computations: algorithms and error correction},}\ }\href@noop {} {\bibfield
  {journal} {\bibinfo  {journal} {Russian Math. Surveys}\ }\textbf {\bibinfo
  {volume} {52}},\ \bibinfo {pages} {1191--1249} (\bibinfo {year}
  {1997})}\BibitemShut {NoStop}%
\bibitem [{\citenamefont {Gottesman}(1997)}]{Gottesman1997}%
  \BibitemOpen
  \bibfield  {author} {\bibinfo {author} {\bibfnamefont {D.}~\bibnamefont
  {Gottesman}},\ }\emph {\bibinfo {title} {Stabilizer codes and quantum error
  correction}},\ \href@noop {} {Ph.D. thesis},\ \bibinfo  {school} {California
  Institute of Technology} (\bibinfo {year} {1997}),\ \bibinfo {note}
  {{\href{https://arxiv.org/abs/quant-ph/9705052}{arXiv:quant-ph/9705052}}}\BibitemShut
  {NoStop}%
\bibitem [{\citenamefont {Knill}\ \emph {et~al.}(1998)\citenamefont {Knill},
  \citenamefont {Laflamme},\ and\ \citenamefont {Zurek}}]{Knill1998}%
  \BibitemOpen
  \bibfield  {author} {\bibinfo {author} {\bibfnamefont {Emanuel}\ \bibnamefont
  {Knill}}, \bibinfo {author} {\bibfnamefont {Raymond}\ \bibnamefont
  {Laflamme}}, \ and\ \bibinfo {author} {\bibfnamefont {Wojciech~H.}\
  \bibnamefont {Zurek}},\ }\bibfield  {title} {\enquote {\bibinfo {title}
  {Resilient quantum computation: error models and thresholds},}\ }\href
  {\doibase 10.1098/rspa.1998.0166} {\bibfield  {journal} {\bibinfo  {journal}
  {Proceedings of the Royal Society of London. Series A: Mathematical, Physical
  and Engineering Sciences}\ }\textbf {\bibinfo {volume} {454}},\ \bibinfo
  {pages} {365--384} (\bibinfo {year} {1998})}\BibitemShut {NoStop}%
\bibitem [{\citenamefont {Preskill}(1998)}]{Preskill1998}%
  \BibitemOpen
  \bibfield  {author} {\bibinfo {author} {\bibfnamefont {John}\ \bibnamefont
  {Preskill}},\ }\bibfield  {title} {\enquote {\bibinfo {title} {Reliable
  quantum computers},}\ }\href {http://www.jstor.org/stable/53172} {\bibfield
  {journal} {\bibinfo  {journal} {Proceedings: Mathematical, Physical and
  Engineering Sciences}\ }\textbf {\bibinfo {volume} {454}},\ \bibinfo {pages}
  {385--410} (\bibinfo {year} {1998})}\BibitemShut {NoStop}%
\bibitem [{\citenamefont {Aliferis}\ \emph {et~al.}(2006)\citenamefont
  {Aliferis}, \citenamefont {Gottesman},\ and\ \citenamefont
  {Preskill}}]{Aliferis2006}%
  \BibitemOpen
  \bibfield  {author} {\bibinfo {author} {\bibfnamefont {Panos}\ \bibnamefont
  {Aliferis}}, \bibinfo {author} {\bibfnamefont {Daniel}\ \bibnamefont
  {Gottesman}}, \ and\ \bibinfo {author} {\bibfnamefont {John}\ \bibnamefont
  {Preskill}},\ }\bibfield  {title} {\enquote {\bibinfo {title} {Quantum
  accuracy threshold for concatenated distance-3 codes},}\ }\href@noop {}
  {\bibfield  {journal} {\bibinfo  {journal} {Quantum Info. Comput.}\ }\textbf
  {\bibinfo {volume} {6}},\ \bibinfo {pages} {97–165} (\bibinfo {year}
  {2006})},\ \bibinfo {note}
  {{\href{https://arxiv.org/abs/quant-ph/0504218}{arXiv:quant-ph/0504218}}}\BibitemShut
  {NoStop}%
\bibitem [{\citenamefont {Aharonov}\ and\ \citenamefont
  {Ben-Or}(2008)}]{Aharonov2008Jul}%
  \BibitemOpen
  \bibfield  {author} {\bibinfo {author} {\bibfnamefont {Dorit}\ \bibnamefont
  {Aharonov}}\ and\ \bibinfo {author} {\bibfnamefont {Michael}\ \bibnamefont
  {Ben-Or}},\ }\bibfield  {title} {\enquote {\bibinfo {title} {{Fault-Tolerant
  Quantum Computation with Constant Error Rate}},}\ }\href {\doibase
  10.1137/S0097539799359385} {\bibfield  {journal} {\bibinfo  {journal} {SIAM
  J. Comput.}\ }\textbf {\bibinfo {volume} {38}},\ \bibinfo {pages}
  {1207--1282} (\bibinfo {year} {2008})}\BibitemShut {NoStop}%
\bibitem [{\citenamefont {Gottesman}(2010)}]{Gottesman2010review}%
  \BibitemOpen
  \bibfield  {author} {\bibinfo {author} {\bibfnamefont {D}~\bibnamefont
  {Gottesman}},\ }\bibfield  {title} {\enquote {\bibinfo {title} {{An
  introduction to quantum error correction and fault-tolerant quantum
  computation}},}\ }in\ \href@noop {} {\emph {\bibinfo {booktitle} {Proceedings
  of Symposia in Applied Mathematics: Quantum Information Science and Its
  Contributions to Mathematics}}},\ Vol.~\bibinfo {volume} {68},\ \bibinfo
  {editor} {edited by\ \bibinfo {editor} {\bibfnamefont {Jr.}\ \bibnamefont
  {Samuel J.~Lomonaco}}}\ (\bibinfo  {publisher} {American Mathematical
  Society},\ \bibinfo {year} {2010})\ pp.\ \bibinfo {pages} {13--58},\ \bibinfo
  {note} {{\href{https://arxiv.org/abs/0904.2557}{Eprint
  arXiv:0904.2557}}}\BibitemShut {NoStop}%
\bibitem [{\citenamefont {Nielsen}\ and\ \citenamefont
  {Chuang}(2010)}]{NielsenChuang2010}%
  \BibitemOpen
  \bibfield  {author} {\bibinfo {author} {\bibfnamefont {Michael~A.}\
  \bibnamefont {Nielsen}}\ and\ \bibinfo {author} {\bibfnamefont {Isaac~L.}\
  \bibnamefont {Chuang}},\ }\href {\doibase 10.1017/CBO9780511976667} {\emph
  {\bibinfo {title} {Quantum Computation and Quantum Information: 10th
  Anniversary Edition}}}\ (\bibinfo  {publisher} {Cambridge University Press},\
  \bibinfo {year} {2010})\BibitemShut {NoStop}%
\bibitem [{\citenamefont {Raussendorf}(2012)}]{Raussendorf2012Sep}%
  \BibitemOpen
  \bibfield  {author} {\bibinfo {author} {\bibfnamefont {Robert}\ \bibnamefont
  {Raussendorf}},\ }\bibfield  {title} {\enquote {\bibinfo {title} {{Key ideas
  in quantum error correction}},}\ }\href {\doibase 10.1098/rsta.2011.0494}
  {\bibfield  {journal} {\bibinfo  {journal} {Philos. Trans. Royal Soc. A}\
  }\textbf {\bibinfo {volume} {370}},\ \bibinfo {pages} {4541--4565} (\bibinfo
  {year} {2012})}\BibitemShut {NoStop}%
\bibitem [{\citenamefont {Devitt}\ \emph {et~al.}(2013)\citenamefont {Devitt},
  \citenamefont {Munro},\ and\ \citenamefont {Nemoto}}]{Devitt2013Jun}%
  \BibitemOpen
  \bibfield  {author} {\bibinfo {author} {\bibfnamefont {Simon~J.}\
  \bibnamefont {Devitt}}, \bibinfo {author} {\bibfnamefont {William~J.}\
  \bibnamefont {Munro}}, \ and\ \bibinfo {author} {\bibfnamefont {Kae}\
  \bibnamefont {Nemoto}},\ }\bibfield  {title} {\enquote {\bibinfo {title}
  {{Quantum error correction for beginners}},}\ }\href {\doibase
  10.1088/0034-4885/76/7/076001} {\bibfield  {journal} {\bibinfo  {journal}
  {Rep. Prog. Phys.}\ }\textbf {\bibinfo {volume} {76}},\ \bibinfo {pages}
  {076001} (\bibinfo {year} {2013})}\BibitemShut {NoStop}%
\bibitem [{\citenamefont {Tuckett}\ \emph {et~al.}(2020)\citenamefont
  {Tuckett}, \citenamefont {Bartlett}, \citenamefont {Flammia},\ and\
  \citenamefont {Brown}}]{Tuckett2020Mar}%
  \BibitemOpen
  \bibfield  {author} {\bibinfo {author} {\bibfnamefont {David~K.}\
  \bibnamefont {Tuckett}}, \bibinfo {author} {\bibfnamefont {Stephen~D.}\
  \bibnamefont {Bartlett}}, \bibinfo {author} {\bibfnamefont {Steven~T.}\
  \bibnamefont {Flammia}}, \ and\ \bibinfo {author} {\bibfnamefont
  {Benjamin~J.}\ \bibnamefont {Brown}},\ }\bibfield  {title} {\enquote
  {\bibinfo {title} {{Fault-Tolerant Thresholds for the Surface Code in Excess
  of 5\% under Biased Noise}},}\ }\href {\doibase
  10.1103/PhysRevLett.124.130501} {\bibfield  {journal} {\bibinfo  {journal}
  {Phys. Rev. Lett.}\ }\textbf {\bibinfo {volume} {124}},\ \bibinfo {pages}
  {130501} (\bibinfo {year} {2020})}\BibitemShut {NoStop}%
\bibitem [{\citenamefont {Brown}(2020)}]{Brown2020May}%
  \BibitemOpen
  \bibfield  {author} {\bibinfo {author} {\bibfnamefont {Benjamin~J.}\
  \bibnamefont {Brown}},\ }\bibfield  {title} {\enquote {\bibinfo {title} {{A
  fault-tolerant non-Clifford gate for the surface code in two dimensions}},}\
  }\href {\doibase 10.1126/sciadv.aay4929} {\bibfield  {journal} {\bibinfo
  {journal} {Sci. Adv.}\ }\textbf {\bibinfo {volume} {6}},\ \bibinfo {pages}
  {eaay4929} (\bibinfo {year} {2020})}\BibitemShut {NoStop}%
\bibitem [{\citenamefont {Bonilla~Ataides}\ \emph {et~al.}(2021)\citenamefont
  {Bonilla~Ataides}, \citenamefont {Tuckett}, \citenamefont {Bartlett},
  \citenamefont {Flammia},\ and\ \citenamefont
  {Brown}}]{BonillaAtaides2021Apr}%
  \BibitemOpen
  \bibfield  {author} {\bibinfo {author} {\bibfnamefont {J.~Pablo}\
  \bibnamefont {Bonilla~Ataides}}, \bibinfo {author} {\bibfnamefont {David~K.}\
  \bibnamefont {Tuckett}}, \bibinfo {author} {\bibfnamefont {Stephen~D.}\
  \bibnamefont {Bartlett}}, \bibinfo {author} {\bibfnamefont {Steven~T.}\
  \bibnamefont {Flammia}}, \ and\ \bibinfo {author} {\bibfnamefont
  {Benjamin~J.}\ \bibnamefont {Brown}},\ }\bibfield  {title} {\enquote
  {\bibinfo {title} {{The XZZX surface code}},}\ }\href {\doibase
  10.1038/s41467-021-22274-1} {\bibfield  {journal} {\bibinfo  {journal} {Nat.
  Commun.}\ }\textbf {\bibinfo {volume} {12}},\ \bibinfo {pages} {1--12}
  (\bibinfo {year} {2021})}\BibitemShut {NoStop}%
\bibitem [{\citenamefont {Fowler}\ \emph {et~al.}(2012)\citenamefont {Fowler},
  \citenamefont {Mariantoni}, \citenamefont {Martinis},\ and\ \citenamefont
  {Cleland}}]{Fowler2012}%
  \BibitemOpen
  \bibfield  {author} {\bibinfo {author} {\bibfnamefont {Austin~G.}\
  \bibnamefont {Fowler}}, \bibinfo {author} {\bibfnamefont {Matteo}\
  \bibnamefont {Mariantoni}}, \bibinfo {author} {\bibfnamefont {John~M.}\
  \bibnamefont {Martinis}}, \ and\ \bibinfo {author} {\bibfnamefont
  {Andrew~N.}\ \bibnamefont {Cleland}},\ }\bibfield  {title} {\enquote
  {\bibinfo {title} {Surface codes: Towards practical large-scale quantum
  computation},}\ }\href {\doibase 10.1103/PhysRevA.86.032324} {\bibfield
  {journal} {\bibinfo  {journal} {Phys. Rev. A}\ }\textbf {\bibinfo {volume}
  {86}},\ \bibinfo {pages} {032324} (\bibinfo {year} {2012})}\BibitemShut
  {NoStop}%
\bibitem [{\citenamefont {Monroe}\ and\ \citenamefont
  {Kim}(2013)}]{Monroe2013}%
  \BibitemOpen
  \bibfield  {author} {\bibinfo {author} {\bibfnamefont {C.}~\bibnamefont
  {Monroe}}\ and\ \bibinfo {author} {\bibfnamefont {J.}~\bibnamefont {Kim}},\
  }\bibfield  {title} {\enquote {\bibinfo {title} {Scaling the ion trap quantum
  processor},}\ }\href {\doibase 10.1126/science.1231298} {\bibfield  {journal}
  {\bibinfo  {journal} {Science}\ }\textbf {\bibinfo {volume} {339}},\ \bibinfo
  {pages} {1164--1169} (\bibinfo {year} {2013})}\BibitemShut {NoStop}%
\bibitem [{\citenamefont {Ikonen}\ \emph {et~al.}(2017)\citenamefont {Ikonen},
  \citenamefont {Salmilehto}, \citenamefont {J.},\ and\ \citenamefont
  {M\"ott\"onen}}]{Miko}%
  \BibitemOpen
  \bibfield  {author} {\bibinfo {author} {\bibfnamefont {J.}~\bibnamefont
  {Ikonen}}, \bibinfo {author} {\bibnamefont {Salmilehto}}, \bibinfo {author}
  {\bibnamefont {J.}}, \ and\ \bibinfo {author} {\bibfnamefont
  {M.}~\bibnamefont {M\"ott\"onen}},\ }\bibfield  {title} {\enquote {\bibinfo
  {title} {Energy-efficient quantum computing},}\ }\href {\doibase
  10.1038/s41534-017-0015-5} {\bibfield  {journal} {\bibinfo  {journal} {npj
  Quantum Inf}\ }\textbf {\bibinfo {volume} {3}},\ \bibinfo {pages} {17}
  (\bibinfo {year} {2017})}\BibitemShut {NoStop}%
\bibitem [{\citenamefont {Terhal}\ and\ \citenamefont
  {Burkard}(2005)}]{Terhal2005Jan}%
  \BibitemOpen
  \bibfield  {author} {\bibinfo {author} {\bibfnamefont {Barbara~M.}\
  \bibnamefont {Terhal}}\ and\ \bibinfo {author} {\bibfnamefont {Guido}\
  \bibnamefont {Burkard}},\ }\bibfield  {title} {\enquote {\bibinfo {title}
  {{Fault-tolerant quantum computation for local non-Markovian noise}},}\
  }\href {\doibase 10.1103/PhysRevA.71.012336} {\bibfield  {journal} {\bibinfo
  {journal} {Phys. Rev. A}\ }\textbf {\bibinfo {volume} {71}},\ \bibinfo
  {pages} {012336} (\bibinfo {year} {2005})}\BibitemShut {NoStop}%
\bibitem [{\citenamefont {Aharonov}\ \emph {et~al.}(2006)\citenamefont
  {Aharonov}, \citenamefont {Kitaev},\ and\ \citenamefont
  {Preskill}}]{Aharonov2006Feb}%
  \BibitemOpen
  \bibfield  {author} {\bibinfo {author} {\bibfnamefont {Dorit}\ \bibnamefont
  {Aharonov}}, \bibinfo {author} {\bibfnamefont {Alexei}\ \bibnamefont
  {Kitaev}}, \ and\ \bibinfo {author} {\bibfnamefont {John}\ \bibnamefont
  {Preskill}},\ }\bibfield  {title} {\enquote {\bibinfo {title}
  {{Fault-Tolerant Quantum Computation with Long-Range Correlated Noise}},}\
  }\href {\doibase 10.1103/PhysRevLett.96.050504} {\bibfield  {journal}
  {\bibinfo  {journal} {Phys. Rev. Lett.}\ }\textbf {\bibinfo {volume} {96}},\
  \bibinfo {pages} {050504} (\bibinfo {year} {2006})}\BibitemShut {NoStop}%
\bibitem [{\citenamefont {Novais}\ and\ \citenamefont
  {Baranger}(2006)}]{Novais2006Jul}%
  \BibitemOpen
  \bibfield  {author} {\bibinfo {author} {\bibfnamefont {E.}~\bibnamefont
  {Novais}}\ and\ \bibinfo {author} {\bibfnamefont {Harold~U.}\ \bibnamefont
  {Baranger}},\ }\bibfield  {title} {\enquote {\bibinfo {title} {{Decoherence
  by Correlated Noise and Quantum Error Correction}},}\ }\href {\doibase
  10.1103/PhysRevLett.97.040501} {\bibfield  {journal} {\bibinfo  {journal}
  {Phys. Rev. Lett.}\ }\textbf {\bibinfo {volume} {97}},\ \bibinfo {pages}
  {040501} (\bibinfo {year} {2006})}\BibitemShut {NoStop}%
\bibitem [{\citenamefont {Novais}\ \emph {et~al.}(2008)\citenamefont {Novais},
  \citenamefont {Mucciolo},\ and\ \citenamefont {Baranger}}]{Novais2008Jul}%
  \BibitemOpen
  \bibfield  {author} {\bibinfo {author} {\bibfnamefont {E.}~\bibnamefont
  {Novais}}, \bibinfo {author} {\bibfnamefont {Eduardo~R.}\ \bibnamefont
  {Mucciolo}}, \ and\ \bibinfo {author} {\bibfnamefont {Harold~U.}\
  \bibnamefont {Baranger}},\ }\bibfield  {title} {\enquote {\bibinfo {title}
  {{Hamiltonian formulation of quantum error correction and correlated noise:
  Effects of syndrome extraction in the long-time limit}},}\ }\href {\doibase
  10.1103/PhysRevA.78.012314} {\bibfield  {journal} {\bibinfo  {journal} {Phys.
  Rev. A}\ }\textbf {\bibinfo {volume} {78}},\ \bibinfo {pages} {012314}
  (\bibinfo {year} {2008})}\BibitemShut {NoStop}%
\bibitem [{\citenamefont {Aliferis}\ and\ \citenamefont
  {Preskill}(2008)}]{Aliferis2008Nov}%
  \BibitemOpen
  \bibfield  {author} {\bibinfo {author} {\bibfnamefont {Panos}\ \bibnamefont
  {Aliferis}}\ and\ \bibinfo {author} {\bibfnamefont {John}\ \bibnamefont
  {Preskill}},\ }\bibfield  {title} {\enquote {\bibinfo {title}
  {{Fault-tolerant quantum computation against biased noise}},}\ }\href
  {\doibase 10.1103/PhysRevA.78.052331} {\bibfield  {journal} {\bibinfo
  {journal} {Phys. Rev. A}\ }\textbf {\bibinfo {volume} {78}},\ \bibinfo
  {pages} {052331} (\bibinfo {year} {2008})}\BibitemShut {NoStop}%
\bibitem [{\citenamefont {Ng}\ and\ \citenamefont
  {Preskill}(2009)}]{Ng2009Mar}%
  \BibitemOpen
  \bibfield  {author} {\bibinfo {author} {\bibfnamefont {Hui~Khoon}\
  \bibnamefont {Ng}}\ and\ \bibinfo {author} {\bibfnamefont {John}\
  \bibnamefont {Preskill}},\ }\bibfield  {title} {\enquote {\bibinfo {title}
  {{Fault-tolerant quantum computation versus Gaussian noise}},}\ }\href
  {\doibase 10.1103/PhysRevA.79.032318} {\bibfield  {journal} {\bibinfo
  {journal} {Phys. Rev. A}\ }\textbf {\bibinfo {volume} {79}},\ \bibinfo
  {pages} {032318} (\bibinfo {year} {2009})}\BibitemShut {NoStop}%
\bibitem [{\citenamefont {Preskill}(2013)}]{Preskill2013Mar}%
  \BibitemOpen
  \bibfield  {author} {\bibinfo {author} {\bibfnamefont {John}\ \bibnamefont
  {Preskill}},\ }\bibfield  {title} {\enquote {\bibinfo {title} {{Sufficient
  condition on noise correlations for scalable quantum computing}},}\ }\href
  {\doibase 10.26421/QIC13.3-4-1} {\bibfield  {journal} {\bibinfo  {journal}
  {Quant. Inf. Comput.}\ }\textbf {\bibinfo {volume} {13}},\ \bibinfo {pages}
  {181--194} (\bibinfo {year} {2013})}\BibitemShut {NoStop}%
\bibitem [{\citenamefont {Novais}\ and\ \citenamefont
  {Mucciolo}(2013)}]{Novais2013Jan}%
  \BibitemOpen
  \bibfield  {author} {\bibinfo {author} {\bibfnamefont {E.}~\bibnamefont
  {Novais}}\ and\ \bibinfo {author} {\bibfnamefont {Eduardo~R.}\ \bibnamefont
  {Mucciolo}},\ }\bibfield  {title} {\enquote {\bibinfo {title} {{Surface Code
  Threshold in the Presence of Correlated Errors}},}\ }\href {\doibase
  10.1103/PhysRevLett.110.010502} {\bibfield  {journal} {\bibinfo  {journal}
  {Phys. Rev. Lett.}\ }\textbf {\bibinfo {volume} {110}},\ \bibinfo {pages}
  {010502} (\bibinfo {year} {2013})}\BibitemShut {NoStop}%
\bibitem [{\citenamefont {Fowler}\ and\ \citenamefont
  {Martinis}(2014)}]{Fowler2014Mar}%
  \BibitemOpen
  \bibfield  {author} {\bibinfo {author} {\bibfnamefont {Austin~G.}\
  \bibnamefont {Fowler}}\ and\ \bibinfo {author} {\bibfnamefont {John~M.}\
  \bibnamefont {Martinis}},\ }\bibfield  {title} {\enquote {\bibinfo {title}
  {{Quantifying the effects of local many-qubit errors and nonlocal two-qubit
  errors on the surface code}},}\ }\href {\doibase 10.1103/PhysRevA.89.032316}
  {\bibfield  {journal} {\bibinfo  {journal} {Phys. Rev. A}\ }\textbf {\bibinfo
  {volume} {89}},\ \bibinfo {pages} {032316} (\bibinfo {year}
  {2014})}\BibitemShut {NoStop}%
\bibitem [{\citenamefont {Jouzdani}\ \emph {et~al.}(2014)\citenamefont
  {Jouzdani}, \citenamefont {Novais}, \citenamefont {Tupitsyn},\ and\
  \citenamefont {Mucciolo}}]{Jouzdani2014Oct}%
  \BibitemOpen
  \bibfield  {author} {\bibinfo {author} {\bibfnamefont {Pejman}\ \bibnamefont
  {Jouzdani}}, \bibinfo {author} {\bibfnamefont {E.}~\bibnamefont {Novais}},
  \bibinfo {author} {\bibfnamefont {I.~S.}\ \bibnamefont {Tupitsyn}}, \ and\
  \bibinfo {author} {\bibfnamefont {Eduardo~R.}\ \bibnamefont {Mucciolo}},\
  }\bibfield  {title} {\enquote {\bibinfo {title} {{Fidelity threshold of the
  surface code beyond single-qubit error models}},}\ }\href {\doibase
  10.1103/PhysRevA.90.042315} {\bibfield  {journal} {\bibinfo  {journal} {Phys.
  Rev. A}\ }\textbf {\bibinfo {volume} {90}},\ \bibinfo {pages} {042315}
  (\bibinfo {year} {2014})}\BibitemShut {NoStop}%
\bibitem [{\citenamefont {Gea-Banacloche}(2002)}]{gea2002some}%
  \BibitemOpen
  \bibfield  {author} {\bibinfo {author} {\bibfnamefont {Julio}\ \bibnamefont
  {Gea-Banacloche}},\ }\bibfield  {title} {\enquote {\bibinfo {title} {{Some
  implications of the quantum nature of laser fields for quantum
  computations}},}\ }\href {\doibase 10.1103/PhysRevA.65.022308} {\bibfield
  {journal} {\bibinfo  {journal} {Phys. Rev. A}\ }\textbf {\bibinfo {volume}
  {65}},\ \bibinfo {pages} {022308} (\bibinfo {year} {2002})}\BibitemShut
  {NoStop}%
\bibitem [{\citenamefont {Steane}(1996)}]{Steane1996}%
  \BibitemOpen
  \bibfield  {author} {\bibinfo {author} {\bibfnamefont {A.~M.}\ \bibnamefont
  {Steane}},\ }\bibfield  {title} {\enquote {\bibinfo {title} {Error correcting
  codes in quantum theory},}\ }\href {\doibase 10.1103/PhysRevLett.77.793}
  {\bibfield  {journal} {\bibinfo  {journal} {Phys. Rev. Lett.}\ }\textbf
  {\bibinfo {volume} {77}},\ \bibinfo {pages} {793--797} (\bibinfo {year}
  {1996})}\BibitemShut {NoStop}%
\bibitem [{Note1()}]{Note1}%
  \BibitemOpen
  \bibinfo {note} {Both $A$ and $A'$ are independent of the nature of the
  errors, and so independent of the system size. The integer $A'=291$ is the
  number of physical gate operations in a CNOT's ``Rec'', and $A=575$ is the
  number in its ``exRec''. ''Rec'' is the encoded CNOT plus the following QEC
  box, while ``exRec'' is the encoded CNOT plus the preceding and following QEC
  boxes, see Ref.~\cite {Aliferis2006}.}\BibitemShut {Stop}%
\bibitem [{Note2()}]{Note2}%
  \BibitemOpen
  \bibinfo {note} {One simply requires that the bath spectrum has cut-offs that
  ensure that $||H_{ij}||$ is not divergent.}\BibitemShut {Stop}%
\bibitem [{\citenamefont {Aliferis}(2013)}]{Aliferis2013Sep}%
  \BibitemOpen
  \bibfield  {author} {\bibinfo {author} {\bibfnamefont {Panos}\ \bibnamefont
  {Aliferis}},\ }\bibfield  {title} {\enquote {\bibinfo {title} {{Introduction
  to quantum fault tolerance}},}\ }in\ \href {\doibase
  10.1017/CBO9781139034807.007} {\emph {\bibinfo {booktitle} {{Quantum Error
  Correction}}}}\ (\bibinfo  {publisher} {Cambridge University Press},\
  \bibinfo {address} {Cambridge, England, UK},\ \bibinfo {year} {2013})\ pp.\
  \bibinfo {pages} {126--160},\ \bibinfo {note} {{Section 5.2.3.1 reviews the
  meaning of the error strength for local non-Markovian noise, a similar
  argument applies for long-range correlations
  \cite{Aharonov2006Feb}.}}\BibitemShut {Stop}%
\bibitem [{Note3()}]{Note3}%
  \BibitemOpen
  \bibinfo {note} {More generally, volume constraints will naturally occur for
  any technology requiring two-qubit gates between arbitrary qubits, if such
  gates become inaccurate at long distances.}\BibitemShut {Stop}%
\bibitem [{Note4()}]{Note4}%
  \BibitemOpen
  \bibinfo {note} {This mechanism would also mean that single qubit gates cause
  local noise on other qubits with nearby frequencies. This has an effect like
  in Sec.~\ref {subsec:ResConstr}, so we do not consider it further here,
  and focus on the effect of $H_{ij}$.}\BibitemShut {Stop}%
\bibitem [{Note5()}]{Note5}%
  \BibitemOpen
  \bibinfo {note} {For example, if driving signals do not affect all qubits
  equally, machine learning can be used to choose each qubit frequency to
  minimizes crosstalk \cite {Klimov2020Jun}, thereby reducing $\beta
  $.}\BibitemShut {Stop}%
\bibitem [{\citenamefont {{M. Fellous-Asiani {\it et al}}}(in
  preparation)}]{our-fullstack}%
  \BibitemOpen
  \bibfield  {author} {\bibinfo {author} {\bibnamefont {{M. Fellous-Asiani {\it
  et al}}}},\ }\href@noop {} {\  (\bibinfo {year} {in
  preparation})}\BibitemShut {NoStop}%
\bibitem [{\citenamefont {Krantz}\ \emph {et~al.}(2019)\citenamefont {Krantz},
  \citenamefont {Kjaergaard}, \citenamefont {Yan}, \citenamefont {Orlando},
  \citenamefont {Gustavsson},\ and\ \citenamefont
  {Oliver}}]{krantz2019quantum}%
  \BibitemOpen
  \bibfield  {author} {\bibinfo {author} {\bibfnamefont {P.}~\bibnamefont
  {Krantz}}, \bibinfo {author} {\bibfnamefont {M.}~\bibnamefont {Kjaergaard}},
  \bibinfo {author} {\bibfnamefont {F.}~\bibnamefont {Yan}}, \bibinfo {author}
  {\bibfnamefont {T.~P.}\ \bibnamefont {Orlando}}, \bibinfo {author}
  {\bibfnamefont {S.}~\bibnamefont {Gustavsson}}, \ and\ \bibinfo {author}
  {\bibfnamefont {W.~D.}\ \bibnamefont {Oliver}},\ }\bibfield  {title}
  {\enquote {\bibinfo {title} {{A quantum engineer's guide to superconducting
  qubits}},}\ }\href {\doibase 10.1063/1.5089550} {\bibfield  {journal}
  {\bibinfo  {journal} {Appl. Phys. Rev.}\ }\textbf {\bibinfo {volume} {6}},\
  \bibinfo {pages} {021318} (\bibinfo {year} {2019})}\BibitemShut {NoStop}%
\bibitem [{\citenamefont {Peropadre}\ \emph {et~al.}(2013)\citenamefont
  {Peropadre}, \citenamefont {Lindkvist}, \citenamefont {Hoi}, \citenamefont
  {Wilson}, \citenamefont {Garcia-Ripoll}, \citenamefont {Delsing},\ and\
  \citenamefont {Johansson}}]{peropadre2013scattering}%
  \BibitemOpen
  \bibfield  {author} {\bibinfo {author} {\bibfnamefont {B.}~\bibnamefont
  {Peropadre}}, \bibinfo {author} {\bibfnamefont {J.}~\bibnamefont
  {Lindkvist}}, \bibinfo {author} {\bibfnamefont {I.-C.}\ \bibnamefont {Hoi}},
  \bibinfo {author} {\bibfnamefont {C.~M.}\ \bibnamefont {Wilson}}, \bibinfo
  {author} {\bibfnamefont {J.~J.}\ \bibnamefont {Garcia-Ripoll}}, \bibinfo
  {author} {\bibfnamefont {P.}~\bibnamefont {Delsing}}, \ and\ \bibinfo
  {author} {\bibfnamefont {G.}~\bibnamefont {Johansson}},\ }\bibfield  {title}
  {\enquote {\bibinfo {title} {{Scattering of coherent states on a single
  artificial atom}},}\ }\href {\doibase 10.1088/1367-2630/15/3/035009}
  {\bibfield  {journal} {\bibinfo  {journal} {New J. Phys.}\ }\textbf {\bibinfo
  {volume} {15}},\ \bibinfo {pages} {035009} (\bibinfo {year}
  {2013})}\BibitemShut {NoStop}%
\bibitem [{\citenamefont {Lodahl}\ \emph {et~al.}(2015)\citenamefont {Lodahl},
  \citenamefont {Mahmoodian},\ and\ \citenamefont
  {Stobbe}}]{lodahl2015interfacing}%
  \BibitemOpen
  \bibfield  {author} {\bibinfo {author} {\bibfnamefont {Peter}\ \bibnamefont
  {Lodahl}}, \bibinfo {author} {\bibfnamefont {Sahand}\ \bibnamefont
  {Mahmoodian}}, \ and\ \bibinfo {author} {\bibfnamefont {S{\o}ren}\
  \bibnamefont {Stobbe}},\ }\bibfield  {title} {\enquote {\bibinfo {title}
  {{Interfacing single photons and single quantum dots with photonic
  nanostructures}},}\ }\href {\doibase 10.1103/RevModPhys.87.347} {\bibfield
  {journal} {\bibinfo  {journal} {Rev. Mod. Phys.}\ }\textbf {\bibinfo {volume}
  {87}},\ \bibinfo {pages} {347--400} (\bibinfo {year} {2015})}\BibitemShut
  {NoStop}%
\bibitem [{\citenamefont {Reiserer}\ and\ \citenamefont
  {Rempe}(2015)}]{Review-Rempe}%
  \BibitemOpen
  \bibfield  {author} {\bibinfo {author} {\bibfnamefont {Andreas}\ \bibnamefont
  {Reiserer}}\ and\ \bibinfo {author} {\bibfnamefont {Gerhard}\ \bibnamefont
  {Rempe}},\ }\bibfield  {title} {\enquote {\bibinfo {title} {Cavity-based
  quantum networks with single atoms and optical photons},}\ }\href {\doibase
  10.1103/RevModPhys.87.1379} {\bibfield  {journal} {\bibinfo  {journal} {Rev.
  Mod. Phys.}\ }\textbf {\bibinfo {volume} {87}},\ \bibinfo {pages}
  {1379--1418} (\bibinfo {year} {2015})}\BibitemShut {NoStop}%
\bibitem [{\citenamefont {Shor}(1994)}]{Shor94}%
  \BibitemOpen
  \bibfield  {author} {\bibinfo {author} {\bibfnamefont {P.}~\bibnamefont
  {Shor}},\ }\bibfield  {title} {\enquote {\bibinfo {title} {Algorithms for
  quantum computation: discrete logarithms and factoring},}\ }in\ \href
  {\doibase 10.1109/SFCS.1994.365700} {\emph {\bibinfo {booktitle} {2013 IEEE
  54th Annual Symposium on Foundations of Computer Science}}}\ (\bibinfo
  {publisher} {IEEE Computer Society},\ \bibinfo {address} {Los Alamitos, CA,
  USA},\ \bibinfo {year} {1994})\ pp.\ \bibinfo {pages} {124--134}\BibitemShut
  {NoStop}%
\bibitem [{\citenamefont {Kjaergaard}\ \emph {et~al.}(2020)\citenamefont
  {Kjaergaard}, \citenamefont {Schwartz}, \citenamefont
  {Braum{\ifmmode\ddot{u}\else\"{u}\fi}ller}, \citenamefont {Krantz},
  \citenamefont {Wang}, \citenamefont {Gustavsson},\ and\ \citenamefont
  {Oliver}}]{kjaergaard2019superconducting}%
  \BibitemOpen
  \bibfield  {author} {\bibinfo {author} {\bibfnamefont {Morten}\ \bibnamefont
  {Kjaergaard}}, \bibinfo {author} {\bibfnamefont {Mollie~E.}\ \bibnamefont
  {Schwartz}}, \bibinfo {author} {\bibfnamefont {Jochen}\ \bibnamefont
  {Braum{\ifmmode\ddot{u}\else\"{u}\fi}ller}}, \bibinfo {author} {\bibfnamefont
  {Philip}\ \bibnamefont {Krantz}}, \bibinfo {author} {\bibfnamefont {Joel
  I.-J.}\ \bibnamefont {Wang}}, \bibinfo {author} {\bibfnamefont {Simon}\
  \bibnamefont {Gustavsson}}, \ and\ \bibinfo {author} {\bibfnamefont
  {William~D.}\ \bibnamefont {Oliver}},\ }\bibfield  {title} {\enquote
  {\bibinfo {title} {{Superconducting Qubits: Current State of Play}},}\ }\href
  {\doibase 10.1146/annurev-conmatphys-031119-050605} {\bibfield  {journal}
  {\bibinfo  {journal} {Annu. Rev. Condens. Matter Phys.}\ }\textbf {\bibinfo
  {volume} {11}},\ \bibinfo {pages} {369--395} (\bibinfo {year}
  {2020})}\BibitemShut {NoStop}%
\bibitem [{\citenamefont {Sears}(2013)}]{sears2013extending}%
  \BibitemOpen
  \bibfield  {author} {\bibinfo {author} {\bibfnamefont {Adam~Patrick}\
  \bibnamefont {Sears}},\ }\href@noop {} {\emph {\bibinfo {title} {Extending
  Coherence in Superconducting Qubits: from microseconds to milliseconds}}}\
  (\bibinfo  {publisher} {Yale University},\ \bibinfo {year}
  {2013})\BibitemShut {NoStop}%
\bibitem [{Note6()}]{Note6}%
  \BibitemOpen
  \bibinfo {note} {For such schemes, $k$ will be replaced by the scale of the
  error correction (i.e. a measure of the number of physical components in the
  error correction). Then one needs to replace our Eq.~(\ref {eq:pk}) with a
  relation (or numerical estimate) for the scaling of the logical error
  strength with $k$, which is not yet available for some schemes.}\BibitemShut
  {Stop}%
\bibitem [{\citenamefont {Cottet}\ \emph {et~al.}(2017)\citenamefont {Cottet},
  \citenamefont {Jezouin}, \citenamefont {Bretheau}, \citenamefont
  {Campagne-Ibarcq}, \citenamefont {Ficheux}, \citenamefont {Anders},
  \citenamefont {Auff{\ifmmode\grave{e}\else\`{e}\fi}ves}, \citenamefont
  {Azouit}, \citenamefont {Rouchon},\ and\ \citenamefont
  {Huard}}]{cottet2017observing}%
  \BibitemOpen
  \bibfield  {author} {\bibinfo {author} {\bibfnamefont
  {Nathana{\ifmmode\ddot{e}\else\"{e}\fi}l}\ \bibnamefont {Cottet}}, \bibinfo
  {author} {\bibfnamefont {S{\ifmmode\acute{e}\else\'{e}\fi}bastien}\
  \bibnamefont {Jezouin}}, \bibinfo {author} {\bibfnamefont {Landry}\
  \bibnamefont {Bretheau}}, \bibinfo {author} {\bibfnamefont {Philippe}\
  \bibnamefont {Campagne-Ibarcq}}, \bibinfo {author} {\bibfnamefont {Quentin}\
  \bibnamefont {Ficheux}}, \bibinfo {author} {\bibfnamefont {Janet}\
  \bibnamefont {Anders}}, \bibinfo {author} {\bibfnamefont {Alexia}\
  \bibnamefont {Auff{\ifmmode\grave{e}\else\`{e}\fi}ves}}, \bibinfo {author}
  {\bibfnamefont {R{\ifmmode\acute{e}\else\'{e}\fi}mi}\ \bibnamefont {Azouit}},
  \bibinfo {author} {\bibfnamefont {Pierre}\ \bibnamefont {Rouchon}}, \ and\
  \bibinfo {author} {\bibfnamefont {Benjamin}\ \bibnamefont {Huard}},\
  }\bibfield  {title} {\enquote {\bibinfo {title} {{Observing a quantum Maxwell
  demon at work}},}\ }\href {\doibase 10.1073/pnas.1704827114} {\bibfield
  {journal} {\bibinfo  {journal} {Proc. Natl. Acad. Sci. U.S.A.}\ }\textbf
  {\bibinfo {volume} {114}},\ \bibinfo {pages} {7561--7564} (\bibinfo {year}
  {2017})}\BibitemShut {NoStop}%
\end{thebibliography}

%

\end{document}